\documentclass[journal,10pt]{IEEEtran}

\usepackage{cite}
\usepackage{graphicx}
\usepackage[cmex10]{amsmath}
\usepackage{amssymb}
\usepackage{booktabs}
\usepackage{threeparttable}
\usepackage{algorithm}
\usepackage{algpseudocode}
\usepackage{amsfonts}
\usepackage{xcolor}
\usepackage{color}

\usepackage{tabularx} %

\usepackage{dblfloatfix} %

\usepackage{soul}

\usepackage{cancel} %

\usepackage{verbatim}%

\usepackage{makecell}

\usepackage{empheq} %

\usepackage{underoverlap} %

\usepackage{hyperref}
\hypersetup{pdftex,colorlinks=true,allcolors=black}

\usepackage{enumitem} %
\usepackage{calc}

\usepackage{textcomp}
\def\BibTeX{{\rm B\kern-.05em{\sc i\kern-.025em b}\kern-.08em
		T\kern-.1667em\lower.7ex\hbox{E}\kern-.125emX}}

\newtheorem{proposition}{\bf Proposition}

\newcommand{\HF}{hopping frequency}
\newcommand{\HFs}{hopping frequencies}

\newcommand{\mj}{\mathsf{j}}

\newcommand{\mn}{\mathsf{r}}

\newcommand{\mb}{\beta}
\newcommand{\ma}{\alpha}

\begin{document}

\title{Integrating Secure and High-Speed Communications into Frequency Hopping MIMO Radar}

\author{
	Kai Wu,
	J. Andrew Zhang,~\IEEEmembership{Senior Member,~IEEE},
	Xiaojing Huang,~\IEEEmembership{Senior Member,~IEEE}, and\\
	Y. Jay Guo,~\IEEEmembership{Fellow,~IEEE}

\thanks{K. Wu, J. A. Zhang, X. Huang and Y. J. Guo are with the Global Big Data Technologies Centre, University of Technology Sydney, Sydney, NSW 2007, Australia (e-mail: kai.wu@uts.edu.au; andrew.zhang@uts.edu.au;  xiaojing.huang@uts.edu.au; jay.guo@uts.edu.au).}%
}

{}

\maketitle

\begin{abstract}
Dual-function radar-communication (DFRC) based on frequency hopping (FH) MIMO radar (FH-MIMO DFRC) achieves symbol rate much higher than radar pulse repetition frequency. 
Such DFRC, however, is prone to eavesdropping due to the spatially uniform illumination of FH-MIMO radar.
How to enhance the physical layer security of FH-MIMO DFRC is vital yet unsolved. In this paper, we reveal the potential of using permutations of \HFs~to achieve secure and high-speed FH-MIMO DFRC.
Detecting permutations at a communication user is challenging due to the dependence on spatial angle. 
We propose a series of baseband waveform processing methods which address the challenge specifically for the legitimate user (Bob) and meanwhile scrambles constellations almost omnidirectionally.  
We discover a deterministic sign rule from the signals processed by the proposed methods. Based on the rule, we develop accurate algorithms for information decoding at Bob. 
Confirmed by simulation, our design achieves substantially high physical layer security for FH-MIMO DFRC, improves decoding performance compared with existing designs and reduces mutual interference among radar targets. 
\end{abstract}


\IEEEpeerreviewmaketitle

\section{Introduction}\label{sec: introduction}

There have been increasing demands for systems with
joint communications and radar sensing capabilities, on
vehicular platforms such as unmanned aerial vehicles and smart
cars \cite{choi2016millimeter}. Performing the two functions on one platform by sharing
hardware and signal processing modules can achieve immediate
benefits of reduced cost, size, weight, and better spectral
efficiency \cite{Andrew_Multibeam2019TVT}.
As pointed out in \cite{Bliss_survey2017}, co-designing joint communication and radar sensing can maximize spectral efficiency with the two sub-systems benefiting each other. Targeting at co-design, some researchers optimize dual-function waveform by jointly considering communication and radar sensing performance metrics (e.g., mutual information and achievable rate etc.), leading to an inherent performance trade-off between the two sub-systems \cite{sturm2011waveformJCAS,LiuFan_waveform2018TSP,XinYuan_JCAS_waveform}. Some researchers exploit the ubiquitous wireless communication technologies, e.g., IEEE 802.11p \cite{surender2011uwb}, IEEE 802.11ad \cite{Robert_82011adTVT,muns2019beam,JointRC_AdaptiveWaveform2019TSP_Robert} and mobile communication network \cite{rahman2019framework}, to perform radar sensing. These design \cite{sturm2011waveformJCAS,JCAS_singleAntennaWaveform2016,LiuFan_waveform2018TSP,XinYuan_JCAS_waveform,surender2011uwb,Robert_82011adTVT,muns2019beam,JointRC_AdaptiveWaveform2019TSP_Robert,rahman2019framework}, however, can have constrained radar sensing ability, as compared with using dedicated radar waveforms.

Integrating secondary data communications in existing radar waveform/platforms, referred to as dual-function radar-communication (DFRC) \cite{hassanien2016signaling,zheng2019radar}, puts radar sensing first. MIMO radar has gained popularity in DFRC given its degrees of freedom in both angle and waveform domains \cite{DFRC_SidelobeControl2016TSP,DFRC_SparseArray2019TAES_XianrongWang,DFRC_waveformShuffling2018DSP,DFRC_CSK2018DSP,DFRC_SP_Mag2019Amin_Aboutanios}. 
Conventional modulations, such as phase shift keying (PSK) and amplitude shift keying, are performed in \cite{DFRC_SidelobeControl2016TSP,DFRC_SparseArray2019TAES_XianrongWang} using the sidelobes of beam pattern. 
Non-traditional modulations, such as waveform shuffling \cite{DFRC_waveformShuffling2018DSP} and code shift keying \cite{DFRC_CSK2018DSP}, have also been developed by optimizing MIMO radar waveform. These works \cite{DFRC_SidelobeControl2016TSP,DFRC_SparseArray2019TAES_XianrongWang,DFRC_waveformShuffling2018DSP,DFRC_CSK2018DSP} generally embed one symbol per one or multiple radar pulses; hence the communication symbol rate is limited by the pulse repetition frequency (PRF).

Employing frequency-hopping (FH) based MIMO (FH-MIMO) radar can increase the symbol rate to much higher than radar PRF, since information embedding can be performed on basis of fast-time sub-pulse \cite{DFRC_AmbiguityFunc2018Amin,FH_MIMO_Radar2019_RadarConf,DFRC_FHcodeSel2018}. 
Hereafter, we refer to FH-MIMO radar-based DFRC as \textit{FH-MIMO DFRC}. 
In \cite{DFRC_AmbiguityFunc2018Amin,FH_MIMO_Radar2019_RadarConf}, PSK-based FH-MIMO DFRC is developed by adding PSK phases onto FH-MIMO radar waveform. 
In \cite{DFRC_FHcodeSel2018}, different combinations of \HFs~are used as constellation points and selected per radar hop (aka sub-pulse within a radar pulse) based on information bits to be transmitted, hence referred to as \HF~combination selection (HFCS). 
HFCS decoding can be readily performed by identifying \HFs~in the frequency domain.
As illustrated in \cite{DFRC_FHcodeSel2018}, HFCS greatly increases the data rate of FH-MIMO DFRC compared with PSK \cite{FH_MIMO_Radar2019_RadarConf}.

{However, HFCS-based FH-MIMO DFRC has a low physical layer security which can hinder its effective application. FH-MIMO radar radiates signals uniformly in the spatial region of interest \cite{book_radarWaveform2012Gini}. Thus, any user covered by radar illumination, albeit the angle-of-departure (AoD) with respect to (w.r.t.) radar, can correctly identify \HFs~to eavesdrop HFCS bits. The detection probability of \HFs~is independent of AoD given the same signal-to-noise ratio (SNR) (which will be detailed in Section \ref{subsec: signal model Eve}). In fact, eavesdropper (Eve), generally having high receiver gain and sensitivity \cite{Kai_PLS_LAA}, can correctly identify the \HFs~with a higher probability compared with a legitimate user. }

\begin{figure*}[!t]
	\centerline{\includegraphics[width=170mm]{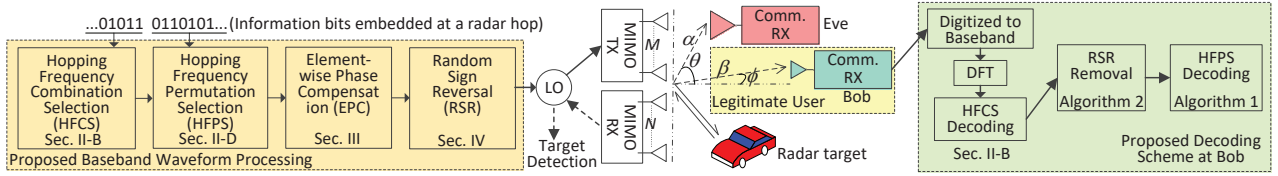}}
	\caption{{System block diagram of an FH-MIMO DFRC, where radar, besides detecting targets, also performs downlink communication through an LoS link with a legitimate user named Bob. Meanwhile, there is an unintended user Eve who eavesdrops on the communication between radar and Bob. The proposed baseband waveform processing highlighted on the left ensures a secure communication by scrambling constellations omnidirectionally. The proposed decoding scheme, as highlighted on the right, can recover constellations at Bob to achieve high-speed data communications.
	}}
	\label{fig: system model}
\end{figure*}

On the other hand, solely using HFCS has not fully exploited the information embedding capability of \HFs. Given any $ M $ \HFs, there are $ M! $ number of permutations, each providing a unique pairing between \HFs~and antennas. 
Thus, in addition to HFCS \cite{DFRC_FHcodeSel2018}, performing \HF~permutation selection (HFPS) at radar and detecting HFPS have the potential of boosting the data rate of FH-MIMO DFRC. 
Unlike HFCS relying on amplitude/power to identify \HFs, HFPS decoding needs to extract signal phases to estimate hopping frequency permutation. This poses a challenging AoD-dependent issue, as will be detailed in Section \ref{subsec: EPC and AOD dependence}.

In this paper, we design new baseband waveform processing to jointly perform HFCS and HFPS in FH-MIMO DFRC, achieving secure and high-speed data communications solely between radar and legitimate user (Bob). We reveal that, besides improving data rate, using HFPS has the substantial potential of enhancing physical layer security for FH-MIMO DFRC.
Our key contributions are summarized as follows.

\begin{enumerate}
	\item Through formulating HFPS decoding problem, we analyze the AoD-dependent issue and accordingly propose an element-wise phase compensation (EPC), removing the AoD dependence of HFPS decoding specifically for Bob. EPC poses a new challenge to Eve by incapacitating HFPS decoding at Eve if not knowing the AoD of Bob;
	
	\item Considering the possible acquisition of the AoD of Bob by Eve, we propose a random sign reversal (RSR) processing which scrambles constellations almost omnidirectionally. We prove that RSR can force the symbol error rate (SER) of Eve into converging to one asymptotically;
	
	\item We discover a deterministic rule related to the signs and phases of the signals processed by EPC and RSR. Based on the rule, we develop an algorithm for Bob to accurately detect and remove RSR. Enabled by EPC, we also design an algorithm for Bob to efficiently decode HFPS. 
\end{enumerate}
We provide a detailed numerical example to demonstrate the overall workflow of incorporating the proposed design in an FH-MIMO DFRC system. We also provide extensive simulations, showing that our design achieves a substantially high communication secrecy and an improved SER performance compared with previous works. 
As also revealed in simulation, the proposed design suppresses sidelobe spikes in the range ambiguity function of FH-MIMO radar, which hence greatly improves signal-to-interference ratio (SIR) of radar detection.
 
\textit{Notations:} The following notations are used throughout the paper. $ C_M^K $ denotes binomial coefficient and $ M! $ denotes $ M $ factorial. $ \lfloor\cdot\rfloor $ rounds towards negative infinity. $ (\cdot)^{\mathrm{T}} $ takes transpose and $ (\cdot)^{\mathrm{*}} $ takes conjugate. $ \|\cdot\|_2 $ denotes $ \ell_2 $-norm. $ [\cdot]_x $ takes element $ x $ of a vector and $ [\cdot]_{x,y} $ takes an element from a matrix at row $ x $ and column $ y $. 
$ \odot $ denotes elementwise product. $ \Re\{x\} $ take the real part of $ x $. 
$ \mathbb{P}\{x=x_0\} $ gives the probability of a random variable $ x $ taking $ x_0 $. 
$ \mathbb{E}\{\cdot\} $ takes expectation.
$ \mathrm{erfc}\{\cdot\} $ denotes the complementary error function.
$ \mathbf{1}_x $ is an $ x $-dimensional unit vector and $ \mathbf{0}_{x\times y} $ denotes an $ x\times y $ matrix of zeros.

\section{Signal Models and Summary of Our Design}\label{sec: signal model}
In this section, the system structure and signal model of 
FH-MIMO DFRC are presented. Fig. \ref{fig: system model} illustrates the overall system block diagram. 
The system consists of an FH-MIMO radar, a single-antenna communication user called Bob and a potential eavesdropper called Eve. The radar is equipped with co-located uniform linear arrays as transmitter and receiver. 
In addition to target detection, the radar also performs downlink data transmission to Bob through an line-of-sight (LoS) channel. 
In this paper, we focus on developing information embedding and decoding schemes to realize secure and high-speed communications between radar and Bob. Thus, we assume that the channel parameters of Bob are available at radar. Also, we consider a practical scenario that the channel information of Eve is unknown to either radar or Bob.

\subsection{FH-MIMO Radar}
Assume that the radar has $ M $ transmitter antennas and $ N $ receiver antennas. Each radar pulse is divided into $ H $ sub-pulses, i.e., hops. Each hop has the time duration of $ T $. The radar frequency band with bandwidth $ B $ is divided evenly into $ K $ sub-bands. The $ k $-th $ (k=0,1,\cdots,K-1) $ sub-band has the baseband central frequency $ kB/K $. 
At hop $ h $ and antenna $ m $, the FH-MIMO radar-transmitted signal is 
\begin{align}\label{eq: transmitted signal p h m}
{s}_{hm}(t) =	e^{\mj2\pi \frac{k_{hm}B}{K} t},~0\le t-h{T} \le T,
\end{align}
where $ k_{hm} $ is the index of the sub-band selected for antenna $ m $ at hop $ h $.
To ensure waveform orthogonality, the following constraints are imposed on radar parameters \cite{Amb_FH_MIMO2008TSP,DFRC_FHcodeSel2018,DFRC_AmbiguityFunc2018Amin}
\begin{align}\label{eq: wavefor orthogonality}
k_{hm}\ne k_{hm'}~\forall m\ne m',~BT/K=\mn(\ge 1),
\end{align}
where $ \mn $ denotes a constant integer.

\subsection{Signal Model of Bob}\label{subsec: signal model for Bob}
Denote the AoD of the LoS path between radar and Bob as $ \phi $ and the complex channel gain of the path as $ \beta $. 
Each hop of signals received at Bob are sampled into $ L $ digital samples by the sampling frequency of $ 2B $. 
Based on (\ref{eq: transmitted signal p h m}), the $ i $-th baseband signal sample received at Bob is given by
\begin{align}\label{eq: comm-received signal}
 y_h(i) 
= \beta \sum_{m=0}^{M-1}e^{-\mj mu_{\phi}} e^{\mj2\pi \mn ik_{hm}/L}  + \xi(i),
\end{align}
where $ u_{\phi}=\frac{2\pi d\sin\phi}{\lambda} $ is referred to as \textit{beamspace AoD} and $ \xi(i) $ is additive white Gaussian noise (AWGN). Here, $ d $ is the antenna spacing of the radar transmitter array and $ \lambda $ is the radar wavelength. 
Taking the $ L $-point DFT of $ y_h(i) $, the result at the $ l $-th discrete frequency is 
\begin{align}\label{eq: DFT of communciation signal}
	& Y_h(l) 
	= \mb\sum_{m=0}^{M-1} e^{-\mj m u_{\phi}} \delta\big(l-\mn k_{hm}\big) + \Xi(l)
\end{align}
where $ \delta(l) $ is the Dirac delta function and $ \Xi(l) $ is the DFT of $ \xi(i) $, i.e., $ \Xi(l)=\frac{1}{L}\sum_{i=0}^{i=L-1}\xi(i)e^{\mj \frac{2\pi li}{L}}~\forall l $. See Appendix \ref{app: derivations of Bob's DFT} for the intermediate calculations of (\ref{eq: DFT of communciation signal}).

According to the waveform orthogonality imposed by (\ref{eq: wavefor orthogonality}), there are $ M $ different \HFs~per hop, which leads to $ M $ non-zero values of the Dirac delta function in (\ref{eq: DFT of communciation signal}). Therefore, by detecting $ M $ peaks in $ |Y_h(l)| $, the \HFs~used at hop $ h $ can be identified. Let $ l_{{m}}^* ~({m}=0,1,\cdots,M-1)$ denote the index of the $ {m} $-th peak of $ |Y_h(l)| $, satisfying
\begin{align}\label{eq: l0<l1<...}
	l_{0}^*<l_1^*<\cdots<l_{M-1}^*.
\end{align}
The $ M $ \HFs~corresponding to the $ M $ peaks can be collected by the following set $ \mathcal{K}_h $, 
\begin{align}\label{eq: Kh}
	\mathcal{K}_h=\left\{ {l_0^*}/{\mn},{l_1^*}/{\mn},\cdots,{l_{M-1}^*}/{\mn} \right\}.
\end{align}
Note that $ \mathcal{K}_h $ is a combination of taking $ M $ out of $ K $ \HFs. The combinations are used as constellation points in \cite{DFRC_FHcodeSel2018}, where the \HFs~per hop are selected at radar based on information bits; and by detecting $ \mathcal{K}_h $ at communication receiver, information is decoded.

\subsection{Signal Model of Eve}\label{subsec: signal model Eve}
As FH-MIMO radar transmits signals omnidirectionally, Eve can receive and process radar signals as Bob does. 
Consider that Eve also takes an $ L $-point DFT per hop.
Let $ \theta $ denote the AoD of Eve, $ \alpha $ the channel gain and $ X_h(l) $ the DFT result at the $ l $-th discrete frequency.
With reference to (\ref{eq: comm-received signal}) and (\ref{eq: DFT of communciation signal}), we can express $ X_h(l) $ as
\begin{align}\label{eq: DFT eave}
	X_h(l) 
	= \ma\sum_{m=0}^{M-1} e^{-\mj m u_{\theta}} \delta\big(l-\mn k_{hm}\big) + Z(l),
\end{align}
where $ u_{\theta}=\frac{2\pi d\sin\theta}{\lambda} $ denotes the beamspace AoD of the LoS path between radar and Eve, and $ Z(l) $ is AWGN at Eve. 

Comparing (\ref{eq: DFT of communciation signal}) and (\ref{eq: DFT eave}), we see that the Dirac delta function takes non-zero values at the same discrete frequencies, i.e., at $ l_m^*~\forall m $ given in (\ref{eq: l0<l1<...}). Without noise, the amplitudes of non-zero $ |Y_h(l)| $ and $ |X_h(l)| $ are $ |\beta| $ and $ |\alpha| $, respectively. Clearly, the power of useful signals received at Bob and Eve are independent of their LoS AoDs w.r.t. radar. That is, $ \mathcal{K}_h $ can be readily identified at Eve as described in Section \ref{subsec: signal model for Bob} for Bob. In fact, due to potentially larger antenna gain \cite{Kai_PLS_LAA}, Eve can have a higher detection probability of $ \mathcal{K}_h $, compared with Bob. This elaborates the low physical layer security of solely using $ \mathcal{K}_h $ for FH-MIMO DFRC, as pointed out in Section \ref{sec: introduction}.

\subsection{Overall Description of Proposed Methods}\label{subsec: overall scheme of the proposed waveform processing}

To achieve a secure and high-speed FH-MIMO DFRC, we
propose new baseband waveform processing, as illustrated in Fig. \ref{fig: system model}.
There are four modules in the proposed processing,
 including HFCS, HFPS, EPC and RSR. Below, we illustrate the first two in details, and then provide the motivations of developing the remaining two modules (which will be presented in subsequent sections).

Besides combinations of \HFs, we also use the permutation of \HFs~to convey information bits. Referring to Fig. \ref{fig: system model}, we divide each communication symbol into two sub-symbols. One is used for HFCS which selects one out of $ C_K^M $ combinations of \HFs~for a radar hop. Given $ M $ \HFs~there are $ M! $ permutations of the frequencies, each providing a pairing between \HFs~and antennas. Thus, we use the second sub-symbol to perform HFPS.

A straightforward benefit of introducing HFPS is the increased data rate given the large number of permutations. Given $ M $ antennas and $ K $ sub-bands, the number of information bits can be conveyed by HFCS and HFPS are up to $  \lfloor \log_2C_M^K \rfloor $ and $  \lfloor \log_2(M!) \rfloor $, respectively. 
Taking $ K=20 $ and $ M=12 $ for example, we have $ \lfloor \log_2C_M^K \rfloor=16 $ and $  \lfloor \log_2(M!) \rfloor=28 $. That is, combining HFCS with HFPS can transmit $ 28 $ more bits per radar hop than solely using HFCS as designed in \cite{DFRC_FHcodeSel2018}.

However, this potential of HFPS is non-trivial to fulfill, due to a challenging AoD-dependent issue in decoding HFPS, as will be illustrated in Section \ref{subsec: EPC and AOD dependence}. To solve the issue, we devise an EPC processing, based on which an algorithm is designed for Bob to efficiently decode HFPS sub-symbol. This will be elaborated on in Section \ref{subsec: EPC and Decoding bob}. 
Due to EPC, HFPS decoding at Eve relies on not only her own AoD but also that of Bob. Therefore, combining HFPS and EPC can potentially enhance the physical layer security of FH-MIMO DFRC, particularly when Eve does not know the AoD of Bob. This will be unveiled in Section \ref{subsec:  EPC on Eve}. 

Considering that a powerful Eve can manage to acquire the AoD of Bob, the secrecy enhanced by EPC is then limited. 
To address this issue, we further propose RSR processing on radar baseband waveform, which severely scrambles constellations almost omnidirectionally. The enhanced secrecy protection against Eve is analyzed in Section \ref{subsec: RSR on Eve}. 
Through investigating the phase changes of the signals processed by EPC and RSR, we discover a deterministic rule to detect and remove RSR for Bob. This will be presented in Section \ref{subsec: RSR at Bob}. In addition, the impact of all the proposed designs on radar detection will be analyzed in Section \ref{subsec: impact on radar}. An exemplary FH-MIMO DFRC integrating our design will be presented in Section \ref{sec: numerical running proposed algorithms}.

\section{Element-wise Phase Compensation}\label{sec: EPC}
In this section, the proposed EPC is presented. Through the elaboration of the AoD-dependent issue in HFPS decoding, we develop EPC processing. Then, the potential of using EPC to enhance physical layer security is unveiled.

\subsection{AoD-Dependence Issue of HFPS Decoding}\label{subsec: EPC and AOD dependence}
To show the AoD dependence of detecting HFPS, i.e., $ \mathbf{k}_h $, we first formulate the detection problem. 
By multiplying $ Y_h(l) $ given in (\ref{eq: DFT of communciation signal}) with $ \mb^* $, the remaining exponential term $ e^{-\mj mu_{\phi}} $ attached to the $ l_{{m}}^* $-th peak is the $ m $-th element of the following steering vector 
\begin{align}\label{eq: a phi steerig vector of Bob}
\mathbf{a}_{\phi}=[1,e^{-\mj u_{\phi}},\cdots,e^{-\mj (M-1) u_{\phi}}]^{\mathrm{T}}.
\end{align}
Stacking the $ L $ peaks of $ Y_h(l) $ into a vector, we have
\begin{align}%
\mathbf{y}_h=[Y_h(l_0^*),Y_h(l_1^*),\cdots,Y_h(l_{M-1}^*)]^{\mathrm{T}}.\nonumber
\end{align}
Based on (\ref{eq: DFT of communciation signal}) and (\ref{eq: l0<l1<...}), we see that $ \mathbf{y}_h  $ is a permutation of the $ \mb $-scaled $ \mathbf{a}_{\phi} $, i.e., 
\begin{align}\label{eq: bf y_h}
	\mathbf{y}_h = \mb\mathbf{P}_{h}\mathbf{a}_{\phi} + \mathbf{v},
\end{align}
where $ \mathbf{P}_{h} $ is permutation matrix and $ \mathbf{v} $ collects $ M $ independent noises $ \Xi(l_m^*)~\forall m $. 
Multiplying $ \mathbf{P}_{h}^{\mathrm{T}} $ to both sides of (\ref{eq: bf y_h}) gives
\begin{align}\label{eq: beta a = Py}
\mathbf{P}_{h}^{\mathrm{T}}\mathbf{y}_h = \mb\mathbf{P}_{h}^{\mathrm{T}}\mathbf{P}_{h}\mathbf{a}_{\phi}+\mathbf{P}_{h}^{\mathrm{T}}\mathbf{v}=\mb\mathbf{a}_{\phi} + \mathbf{P}_{h}^{\mathrm{T}}\mathbf{v},
\end{align}
where $ \mathbf{P}_{h}^{\mathrm{T}}\mathbf{P}_{h}=\mathbf{I} $ is due to the orthogonal property of a permutation matrix. 
By comparing the two sides of (\ref{eq: beta a = Py}) in a pointwise manner, the \HFs~attached to the $ M $ antennas can be expressed as
\begin{align}\label{eq: hat k_h}
{\mathbf{k}}_{h} = \mathbf{P}_{h}^{\mathrm{T}} [l^*_{0}/\mn,l^*_{1}/\mn,\cdots,l^*_{M-1}/\mn]^{\mathrm{T}}.
\end{align}  
Eq. (\ref{eq: hat k_h}) indicates that HFPS decoding can be performed by identifying the permutation matrix $ \mathbf{P}_h $.

According to (\ref{eq: beta a = Py}), the following detector can be formulated to identify $ \mathbf{P}_h $, 
\begin{align}\label{eq: ML}
\mathbf{P}_h:~\min_{\mathbf{P}\in \mathcal{P}}  g(\mathbf{P})~~~\mathrm{s.t.}~g(\mathbf{P})=\| \mb  \mathbf{a}_{\phi} - \mathbf{P}^{\mathrm{T}} \mathbf{y}_h \|_2^2,
\end{align}
where $ \mathcal{P} $ is the set of possible permutation matrices. 
From (\ref{eq: ML}), we see that the detecting performance of $ \mathbf{P}_h $ varies with the AoD of Bob, i.e., $ \phi $. 
This leads to the so-called AoD-dependent issue of HFPS decoding. In particular, given a small $ \phi $, the distance between adjacent elements in $ \mathbf{a}_{\phi} $ is also small, leading to a high error probability of detecting $ \mathbf{P}_h $ and decoding HFPS. 
This issue can be relieved with a large $ \phi $. However, a large $ \phi $ can lead to an ambiguity issue. When $ u_{\phi}>\frac{2\pi}{M} $, it can happen that the phases of $ e^{-\mj m u_{\phi}} $ at two or more different values of $ m $ are identical.

 \subsection{EPC and HFPS Decoding at Bob}\label{subsec: EPC and Decoding bob}
We see from (\ref{eq: a phi steerig vector of Bob}) the elements of $ \mathbf{a}_{\phi} $ are similar to the constellation points of the $ M $-PSK modulation. Thus, when $ e^{-\mj mu_{\phi}}~\forall m $ are uniformly distributed on the unit circle, the best detecting performance can be achieved \cite{DFRC_waveformShuffling2018DSP}.
In light of this, we introduce EPC to compensate the phase of each antenna-transmitted signal so that $ \mathbf{a}_{\phi} $ given in (\ref{eq: a phi steerig vector of Bob}) becomes the following AoD-independent steering vector
\begin{align}\label{eq: stv constant}
{\mathbf{a}} = [1,e^{-\mj 2\pi/M},\cdots,e^{-\mj 2\pi(M-1)/M}]^{\mathrm{T}}.
\end{align}
Based on (\ref{eq: transmitted signal p h m}), (\ref{eq: comm-received signal}) and (\ref{eq: DFT of communciation signal}), the following EPC is introduced for the $ m $-th antenna at hop $ h $, 
\begin{align}\label{eq: EPC waveform}
	\tilde{s}_{hm} = s_{hm}(t) e^{-\mj m(2\pi/M-u_{\phi})}.
\end{align}
\textit{We add a tilde sign above the relevant variables to reflect the impact of EPC.} That is, $ y_h(l) $, $ Y_h(l) $, $ \mathbf{y}_h $ and $ g(\mathbf{P}) $, as given in (\ref{eq: comm-received signal}), (\ref{eq: DFT of communciation signal}), (\ref{eq: bf y_h}) and (\ref{eq: ML}), respectively, are now denoted by $ \tilde{y}_h(l) $, $ \tilde{Y}_h(l) $, $ \tilde{\mathbf{y}}_h $ and $ \tilde{g}(\mathbf{P}) $. 
Note that 
\begin{itemize}
	\item $ \tilde{y}_h(l) $ has the same expression as $ y_h(l) $ except that $ u_{\phi} $ is replaced by $ 2\pi/M $. The same goes for $ \tilde{Y}_h(l) $;
	
	\item $ \tilde{\mathbf{y}}_h $ replaces $ \mathbf{a}_{\phi} $ in (\ref{eq: bf y_h}) with $ \mathbf{a} $ given in (\ref{eq: stv constant}). The same permutation matrix is applicable for both $ \tilde{\mathbf{y}}_h $ and $ {\mathbf{y}}_h $, since EPC does not change the indexes of peaks in $ \tilde{Y}_h(l) $; refer to Section \ref{subsec: signal model Eve}.
\end{itemize}
Jointly considering the above changes caused by EPC, the objective function of problem (\ref{eq: ML}) 
becomes $ \tilde{g}(\mathbf{P})=\| \mb  \mathbf{a} - \mathbf{P}^{\mathrm{T}} \tilde{\mathbf{y}}_h \|_2^2 $.
Expanding the objective function $ \tilde{g}(\mathbf{P}) $ gives
\begin{align}
& \tilde{g}(\mathbf{P}) = \|\mb\mathbf{a}\|_2^2 + \|\tilde{\mathbf{y}}_h\|_2^2-2\Re\big\{ \mb^*{\mathbf{a}}^{\mathrm{H}} \mathbf{P}^{\mathrm{T}} \tilde{\mathbf{y}}_h \big\}.\nonumber
\end{align}
We see that minimizing $ \tilde{g}(\mathbf{P}) $ is equivalent to maximizing $ \Re\big\{ \mb^*{\mathbf{a}}^{\mathrm{H}} \mathbf{P}^{\mathrm{T}} \tilde{\mathbf{y}}_h \big\} $ w.r.t. $ \mathbf{P} $, as the other two terms are irrelevant to $ \mathbf{P} $. The maximization is achieved when the pointwise phase difference between $ {\mathbf{a}} $ and $ \mb^* \mathbf{P}^{\mathrm{T}} \tilde{\mathbf{y}}_h $ is minimized. Due to EPC, the element phases of $ \mathbf{a} $ are in descending order for sure.
This implies that the correct $ \mathbf{P} $ needs to sort the element phases of $ \mb^* \mathbf{P}_h^{\mathrm{T}} \tilde{\mathbf{y}}_h $ in descending order as well. Let the $ M\times 1 $ vector $ \mathbf{i} $ collects the arrangements of the element phases of $ \mb^*\tilde{\mathbf{y}}_h $ into the sorted version. We propose the following solution to (\ref{eq: ML}),
\begin{align}\label{eq: Ph solution}
&[\mathbf{P}_h]_{[\mathbf{i}]_m,m}=1~ (m=0,1,\cdots,M-1)
\end{align}
where $ \mathbf{P}_h $ is initialized as $ \mathbf{0}_{M\times M} $.  

Based on the above analysis, Algorithm \ref{alg: detect HFPS} is established to decode HFPS at Bob. In Step \ref{stepA1: angle taking}, directly taking the element angles makes the values in {$ \vec{{\omega}} $ fall in the region of $ [-\pi,\pi] $}. Thus, Step \ref{stepA1: angle revision} recovers true angles by compensating $ -2\pi $ on elements larger than $ 0 $. This is because all the element angles of $ \mb^*\tilde{\mathbf{y}}_h $ should be equal to those of $ \mathbf{a} $ given in (\ref{eq: stv constant}) in the absence of noise, and taking angle in Step \ref{stepA1: angle taking} adds $ 2\pi $ onto any angle smaller than $ -\pi $. Corrupted in noises, the phase of $ [\mb^*\tilde{\mathbf{y}}_h]_0 $ (which is zero without noise) can become a small positive value which will then be revised through Step \ref{stepA1: angle revision} into a small negative value close to $ -2\pi $. This phenomenon, known as zigzag \cite{KaiWu_NrrbndAoA2017TWC}, turns the largest element angle of $ \mb^*\tilde{\mathbf{y}}_h $ into the smallest one.
Step \ref{stepA1: circshift} is introduced to remove potential zigzag by comparing the Euclidean distances $ \|e^{\mj [\vec{\omega}]_{[\mathbf{i}]_{M-1}}}-1\|_2 $ and $ \|e^{\mj [\vec{\omega}]_{[\mathbf{i}]_{0}}}-1\|_2 $. Based on (\ref{eq: hat k_h}) and (\ref{eq: Ph solution}), Steps \ref{stepA1: Ph construction} and \ref{stepA1: bf k_h solving} produce $ \mathbf{k}_h $ which is finally used for HFPS decoding in Step \ref{stepA1: final decoding}.

\begin{algorithm}[!t] %
	
	\caption{HFPS Decoding at Bob}
	\label{alg: detect HFPS}
	\begin{algorithmic}[1]	
		
		\State{\textbf{Input}: $ \tilde{\mathbf{y}}_h $, $ \mb $, $ l_m^*(\forall m) $ and $ \mn $ (given in (\ref{eq: wavefor orthogonality}));} \label{stepA1: input}
		
		\State{Take the element angles of $ \mb^*\tilde{\mathbf{y}}_h $, and stack them $ \vec{\omega} $;}\label{stepA1: angle taking}
		
		\State{At $ \forall m $, if $ [\vec{\omega}]_m>0 $, $ [\vec{\omega}]_m=[\vec{\omega}]_m-2\pi $;}
		\label{stepA1: angle revision}

		\State{Sorting the elements of $ \vec{\omega} $ in descending order gives $ \mathbf{i} $;
		 }\label{stepA1: angle sorting}
	 
	 \State{If $ \|e^{\mj [\vec{\omega}]_{[\mathbf{i}]_{M-1}}}-1\|_2<\|e^{\mj [\vec{\omega}]_{[\mathbf{i}]_{0}}}-1\|_2 $, shift $ \mathbf{i} $ circularly by a single element;}\Comment{{\footnotesize Similar to ``circshift()'' in MATLAB;}}\label{stepA1: circshift}

		\State{Substitute $ \mathbf{i} $ into (\ref{eq: Ph solution}) to construct $ \mathbf{P}_h $;}\label{stepA1: Ph construction}
		
		\State{Substituting $ \mathbf{P}_h $, $ l_m^* $ and $ \mn $ into (\ref{eq: hat k_h}) leads to $ \mathbf{k}_h $;}\label{stepA1: bf k_h solving}
		
		\State{Look up $ \mathbf{k}_h $ in the constellation set to decode HFPS.}		\label{stepA1: final decoding}
	\end{algorithmic}
\end{algorithm}

\subsection{Enhancing Physical Layer Security by EPC} \label{subsec:  EPC on Eve}
As the phase compensation of EPC is determined based on the AoD of Bob, the received signal at Eve is still AoD-dependent. To this end, we can conclude that EPC helps enhance the physical layer security of FH-MIMO DFRC. 
Taking into account EPC in the frequency-domain signal received at Eve, i.e., $ X_h(l) $ given in (\ref{eq: DFT eave}), the signal can be rewritten into
\begin{align}\label{eq: DFT eave EPC}
\tilde{X}_h(l) 
= \ma\sum_{m=0}^{M-1} e^{-\mj m (\frac{2\pi}{M}+u_{\theta}-u_{\phi})} \delta\big(l-\mn k_{hm}\big) + Z(l).
\end{align}
Again, we notice that detecting the set of \HFs, i.e., $ \mathcal{K}_h $, at Eve is not affected by AoD. The same set $ \{l_m^*~\forall m\} $ given in (\ref{eq: Kh}) can be identified at Eve as that obtained at Bob. 
Similar to $ \mathbf{y}_h $ given in (\ref{eq: bf y_h}), the following vector can be obtained at Eve,
\begin{align}\label{eq: tilde bf x_h}
	\tilde{\mathbf{x}}_h = [\tilde{X}_h(l_0^*),\tilde{X}_h(l_1^*),\cdots,\tilde{X}_h(l_{M-1}^*) ]^{\mathrm{T}}.
\end{align}
With reference to (\ref{eq: beta a = Py})$ \sim $(\ref{eq: ML}), the following detector is formulated at Eve to decode HFPS by identifying the permutation matrix $ \mathbf{P}_h $,
\begin{align}\label{eq: ML eave}
	&{\mathbf{P}}_h:\min_{{\mathbf{P}}\in \mathcal{P}_{\mathrm{Eve}}}  {f}({\mathbf{P}})\\
	\text{ s.t. } ~&{f}({\mathbf{P}}) =
		\left\{ 
		\begin{array}{ll}
			\| \ma  {\mathbf{a}}_{\theta\phi}  - {\mathbf{P}}^{\mathrm{T}} \tilde{\mathbf{x}}_h \|_2^2, &\text{ if Eve knows }u_{\phi}\\
			\| \ma  {\mathbf{a}}_{\theta}  - {\mathbf{P}}^{\mathrm{T}} \tilde{\mathbf{x}}_h \|_2^2, &\text{ otherwise}
		\end{array}
		\right.,\nonumber
\end{align}
where the two steering vectors are 
\begin{gather}
	{\mathbf{a}}_{\theta\phi}=[1,e^{-\mj  (\frac{2\pi}{M}+u_{\theta}-u_{\phi})},\cdots,e^{-\mj (M-1) (\frac{2\pi}{M}+u_{\theta}-u_{\phi})}]^{\mathrm{T}}\text{ and }\nonumber\\
	{\mathbf{a}}_{\theta}=[1,e^{-\mj  (\frac{2\pi}{M}+u_{\theta})},\cdots,e^{-\mj (M-1) (\frac{2\pi}{M}+u_{\theta})}]^{\mathrm{T}}.\nonumber
\end{gather}
Note that the same $ \mathbf{P}_h $ is required for Eve and Bob to decode HFPS, since $ \mathbf{P}_h $ is added at radar as part of the waveform.

From (\ref{eq: ML eave}), we see that EPC makes decoding HFPS at Eve relies on not only the AoD of Eve $ u_{\theta} $ but also that of Bob $ u_{\phi} $. If $ u_{\phi} $ is unknown to Eve, the decoding 
performance can degrade drastically, since the actual steering vector contained in $ \tilde{\mathbf{x}}_h $ is $ {\mathbf{a}}_{\theta\phi} $. 
A powerful eavesdropper may manage to know $ u_{\phi} $, reducing the secrecy enhancement brought by EPC.
This can be solved by a new technique proposed below.

\section{Random Sign Reversal}\label{sec: RSR}
To further enhance the secrecy of FH-MIMO DFRC, we introduce RSR to scramble the constellations received by Eve, even when the AoD of Bob is known to Eve. As indicated by the name, RSR randomly selects several antennas and reverse the signs of signals transmitted by them. Antennas are randomly and independently selected over hops. In this section, we first explain how RSR can scramble constellations received by Eve, and then develop an algorithm for Bob to remove RSR. We also elaborate the impact of all proposed modules on radar performance. \textit{In the following, we add a breve sign above relevant variables to indicate RSR processing, e.g., $ \breve{s}_{hm}(t) $ corresponding to the EPC-processed $ \tilde{s}_{hm}(t) $ given in (\ref{eq: EPC waveform}) and the original radar waveform $ s_{hm}(t) $ given in (\ref{eq: comm-received signal}).}

\subsection{Impact of RSR on Eve}\label{subsec: RSR on Eve}

We can reflect RSR in radar-transmitted signals by adding a binary coefficient $ b_{hm} $ onto $ \tilde{s}_{hm}(t) $ given in (\ref{eq: EPC waveform}), leading to 
\begin{align}\label{eq: radar signal after RSR}
	\breve{s}_{hm}(t) = b_{hm}\tilde{s}_{hm}(t)=b_{hm}s_{hm}(t) e^{-\mj m(2\pi/M-u_{\phi})}.
\end{align} 
Denoting the number of sign-reversed antennas as $ Q $ per hop, we have
\begin{align}\label{eq: P bhm}
	\mathbb{P}\{b_{hm}=1\}={(M-Q)}/{M};~\mathbb{P}\{b_{hm}=-1\}={Q}/{M},
\end{align}
where $ \mathbb{P} $ takes probability.
Substituting (\ref{eq: radar signal after RSR}) into (\ref{eq: DFT eave EPC}) and (\ref{eq: tilde bf x_h}), we can rewrite $ \tilde{\mathbf{x}}_h $ into 
\begin{align}\label{eq: RSR signal vector eave }
	\breve{\mathbf{x}}_h = \ma{\mathbf{P}}_h\big(\mathbf{b}_h\odot {\mathbf{a}}_{\theta\phi}\big)+{\mathbf{z}},
\end{align}
where $ \odot $ denotes pointwise product and $ {\mathbf{z}} $ collects $ Z(l_m^*)~\forall m $. 

Replacing $ \tilde{\mathbf{x}}_h $ in (\ref{eq: ML eave}) with $ \breve{\mathbf{x}}_h $, the objective function, denoted by $ f_b({\mathbf{P}}) $ (with the subscript added to reflect the impact of RSR), becomes
\begin{align}\label{eq: objective function change of ML eave}
	&f_b({\mathbf{P}})   =  \| \ma  {\mathbf{a}}_{\theta\phi}  - \ma{\mathbf{P}}^{\mathrm{T}}{\mathbf{P}}_h\big(\mathbf{b}_h\odot {\mathbf{a}}_{\theta\phi}\big) - {\mathbf{P}}^{\mathrm{T}} {\mathbf{z}} \|_2^2 \nonumber\\
	& = 2M|\ma|^2 +  \tilde{z}- 2|\ma|^2 \underbrace{\Re\big\{ {\mathbf{a}}_{\theta\phi}^{\mathrm{H}} {\mathbf{P}}^{\mathrm{T}}{\mathbf{P}}_h\big(\mathbf{b}_h\odot {\mathbf{a}}_{\theta\phi}\big)  \big\} }_{{f}_{b}^{\Re}({\mathbf{P}})}
\end{align} 
where $ \| {\mathbf{a}}_{\theta\phi}\|_2^2=M $ and $ \|{\mathbf{P}}^{\mathrm{T}}{\mathbf{P}}_h\big(\mathbf{b}_h\odot {\mathbf{a}}_{\theta\phi}\big)\|_2^2=M $ are plugged in, and $ \tilde{z} $ is the sum of noise-related terms. It can be readily confirmed that the variance of $ \tilde{z} $ is independent of $ {\mathbf{P}} $ and $ \mathbf{b}_h $. 
Without RSR, i.e., $ \mathbf{b}_h=\mathbf{1} $ in (\ref{eq: objective function change of ML eave}), $ {f}_{b}^{\Re}({\mathbf{P}}) $ is maximized at $ \mathbf{P}=\mathbf{P}_h $, which then minimizes $ f_b({\mathbf{P}}) $ without noise. However, affected by RSR, $ f_b({\mathbf{P}}) $ can no longer be maximized at $ \mathbf{P}=\mathbf{P}_h $, as detailed in the following.

\vspace{3pt}
\begin{proposition}\label{pp: fb Gaussian noise and SER convegernce}
	\textit{Within the following angular region
	\begin{align}\label{eq: u_thetaPhi}
	0\le u_{\theta\phi}\le \frac{2\pi(M-2)}{M},~\mathrm{s.t.}~u_{\theta\phi}=u_{\theta}-u_{\phi},
	\end{align}
	the proposed RSR makes $ {f}_{b}^{\Re}({\mathbf{P}}) $ approach a normal distribution with parameters given in (\ref{eq: mean and variance of artificial noise}),
	which further forces the SER of Eve, who solves (\ref{eq: ML eave}) for HFPS decoding, into converging to one in high SNR regions.	
	\begin{align}\label{eq: mean and variance of artificial noise}
		\mu_f =  \frac{M-2Q}{M^2} \frac{\sin^2\left(\frac{M(u_{\theta\phi}+\frac{2\pi}{M})}{2}\right)}{\sin^2\left(\frac{(u_{\theta\phi}+\frac{2\pi}{M})}{2}\right)},\sigma_f^2=\frac{1}{2}(M-\mu_f^2)
	\end{align}}	
\end{proposition}

\begin{figure}[!t]
	\centerline{\includegraphics[width=80mm]{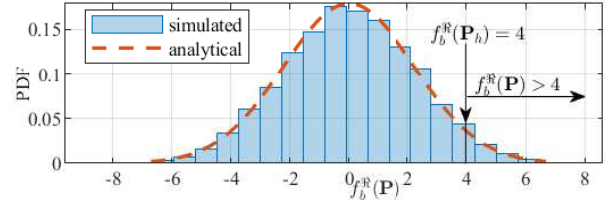}}
	\caption{Histogram of simulated $ {f}_{b}^{\Re}({\mathbf{P}}) $ and the analytical PDF based on the parameters derived in Proposition \ref{pp: fb Gaussian noise and SER convegernce}, where $ u_{\theta\phi}=0 $, $ M=8 $ and $ Q=2 $.}
	\label{fig: RSR Gaussian distribution}
\end{figure}

\begin{figure}[!t]
	\centerline{\includegraphics[width=80mm]{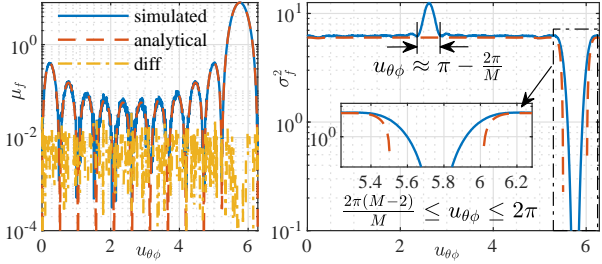}}
	\caption{The mean and variance of $ {f}_{b}^{\Re}({\mathbf{P}}) $ vs $ u_{\theta\phi} $ with $ M=12 $ and $ Q=2 $.}
	\label{fig: RSR mean and var}
\end{figure}

Refer to Appendix \ref{app: fb Gaussian noise} for the proof of Proposition \ref{pp: fb Gaussian noise and SER convegernce}. Fig. \ref{fig: RSR Gaussian distribution} plots the histogram of the simulated 
$ {f}_{b}^{\Re}({\mathbf{P}}) $ with the analytical PDF plotted based on the parameters derived in Proposition \ref{pp: fb Gaussian noise and SER convegernce}. We see that, as proved, $ {f}_{b}^{\Re}({\mathbf{P}}) $ conforms to a normal distribution. We also see that the analytical PDF of $ {f}_{b}^{\Re}({\mathbf{P}}) $ overlaps with the simulated one, which confirms the accuracy of the parameters derived in Proposition \ref{pp: fb Gaussian noise and SER convegernce}. Given $ M=8 $ and $ Q=2 $ in Fig. \ref{fig: RSR Gaussian distribution}, we have $ {f}_{b}^{\Re}({\mathbf{P}}_h)=4 $. Seen from the figure, there are a non-negligible number of $ \mathbf{P}(\ne \mathbf{P}_h) $ leading to $ {f}_{b}^{\Re}({\mathbf{P}})>{f}_{b}^{\Re}({\mathbf{P}}_h) $. In this case, solving (\ref{eq: ML eave}) returns the $ \mathbf{P} $ in the region of $ {f}_{b}^{\Re}({\mathbf{P}})>4 $ (as annotated in Fig. \ref{fig: RSR Gaussian distribution}) rather than $ \mathbf{P}_h $ required by Eve for HFPS decoding.

Fig. \ref{fig: RSR mean and var} compares the simulated and analytical parameters of $ {f}_{b}^{\Re}({\mathbf{P}}) $ against $ u_{\theta\phi} $. We see that the analytical $ \mu_f $ given in (\ref{eq: mean and variance of artificial noise}) depicts the mean of $ {f}_{b}^{\Re}({\mathbf{P}}) $ accurately over the whole angular region, and the analytical $ \sigma_f^2 $ is accurate for most angles. 
An exceptional angular region where the analytical and simulated values of $ \sigma_f^2 $ have a non-negligible gap is around $ u_{\theta\phi}=\pi-\frac{2\pi}{M} $. In this region, the complex $ {f}_{b}^{\Re}({\mathbf{P}}) $ has much smaller imaginary parts (compared with real parts) and even turns into a real value at
$ u_{\theta\phi}=\pi-\frac{2\pi}{M} $; see (\ref{eq: objective function change of ML eave}). 
Since we take half the variance of $ {f}_{b}({\mathbf{P}}) $ as that of its real part $ {f}_{b}^{\Re}({\mathbf{P}}) $ (see (\ref{eq: sigma^2_f}) in Appendix \ref{app: fb Gaussian noise}), the actual variance of $ {f}_{b}^{\Re}({\mathbf{P}}) $ is larger than $ \sigma_f^2 $ given in (\ref{eq: mean and variance of artificial noise}) when $ u_{\theta\phi} $ is close to $ \pi-\frac{2\pi}{M} $ and doubles $ \sigma_f^2 $ at $ u_{\theta\phi}=\pi-\frac{2\pi}{M} $. 
In fact, this exception is favorable to improve communication secrecy. As seen in Fig. \ref{fig: RSR Gaussian distribution}, a larger variance of $ {f}_{b}^{\Re}({\mathbf{P}}) $ increases the number of permutation matrices causing incorrect HFPS decoding at Eve.

Note that, although the region $ \frac{2\pi(M-2)}{M}\le u_{\theta\phi}\le 2\pi $, denoted by $ \mathcal{R} $, is excluded in Proposition \ref{pp: fb Gaussian noise and SER convegernce};
seen from the zoomed-in sub-figure in Fig. \ref{fig: RSR mean and var}, the analytical $ \sigma_f^2 $ derived in the proposition can depict the actual variance in part of $ \mathcal{R} $. (The reason for excluding $ \mathcal{R} $ is illustrated at the end of Appendix \ref{app: fb Gaussian noise}.) The decoding performance of Eve is low in $ \mathcal{R} $, as explained below. We see from Fig. \ref{fig: RSR mean and var} that $ \mathcal{R} $ can be divided into two sub-regions, denoted by $ \mathcal{R}_1 $ and $ \mathcal{R}_2 $, with $ \sigma_f^2> 0 $ and $ \sigma_f^2\approxeq 0 $, respectively. In $ \mathcal{R}_1 $, since $ \sigma_f^2> 0 $, the conclusion in Proposition \ref{pp: fb Gaussian noise and SER convegernce} still holds, i.e., the normally distributed $ f_b^{\Re}(\mathbf{P}) $ leads to the asymptotic convergence of the SER at Eve. In $ \mathcal{R}_2 $, due to $ u_{\theta\phi}+\frac{2\pi}{M}\approxeq 2\pi $, we have $ \mathbf{a}_{\theta\phi}\approxeq\mathbf{1}_M $, which degrades the decoding performance of Eve and even invalidates (\ref{eq: ML eave}) at $ u_{\theta\phi}+\frac{2\pi}{M}= 2\pi $.

\subsection{Detecting RSR at Bob}\label{subsec: RSR at Bob}
Enabled by EPC, Bob can recover RSR-scrambled constellations, which is developed in this subsection. 
Similar to (\ref{eq: RSR signal vector eave }), EPC and RSR turn $ \mathbf{y}_h $ given in (\ref{eq: bf y_h}) into 
\begin{align} \label{eq: bf y_h rewritten}
 \breve{\mathbf{y}}_h 	=\mb\mathbf{P}_h \breve{\mathbf{a}} + \mathbf{v},~\mathrm{s.t.}~\breve{\mathbf{a}}=\big(\mathbf{b}_h\odot\mathbf{a}\big)
\end{align}
where $ \mathbf{a} $ is given in (\ref{eq: stv constant}) and $ \mathbf{v} $ collects the AWGNs $ \Xi(l_{{m}}^*)~\forall{m} $ given in (\ref{eq: DFT of communciation signal}). We discover a deterministic rule concerning the element signs and phases of $ \breve{\mathbf{a}} $ --- \textit{the joint processing of EPC and RSR turns the $ m $-th element into another one in $ \breve{\mathbf{a}} $.}
By assuming $ b_{hm}=-1 $, $ [\breve{\mathbf{a}}]_m $ becomes 
\begin{align}
[\breve{\mathbf{a}}]_m = -e^{-\mj \frac{2\pi m}{M}}=e^{\mj \frac{2\pi M/2}{M}}e^{-\mj \frac{2\pi m}{M}}=[\breve{\mathbf{a}}]_{m\pm\frac{M}{2}},
\end{align}
where $ m\pm\frac{M}{2} $ depends on $ m\lessgtr\frac{M}{2} $.

The discovered rule enables us to detect RSR by identifying identical elements in $ \breve{\mathbf{a}} $.
However, to ensure correct detection of RSR, two constraints are necessary. \textit{First},
when $ b_{hm}=-1 $ and $ b_{h(m+\frac{M}{2})}=-1 $ happen simultaneously, RSR turns $ [\breve{\mathbf{a}}]_m $ and $ [\breve{\mathbf{a}}]_{m+\frac{M}{2}} $ into each other. In this case, two reversed antennas will be identified as a single one. To avoid this, we impose the constraint that $ b_{hm}=-1 $ and $ b_{h(m+\frac{M}{2})}=-1 $ cannot happen simultaneously, i.e.,
\begin{align}\label{eq: C1}
b_{hm}+b_{h(m+{M}/{2})}\ne -2~~ \forall m<M/2.
\end{align}
\textit{Second}, both $ b_{hm}=-1 $ and $ b_{h(m+\frac{M}{2})}=-1 $ lead to $ [\breve{\mathbf{a}}]_m=[\breve{\mathbf{a}}]_{m+\frac{M}{2}} $. 
In turn, $ [\breve{\mathbf{a}}]_m=[\breve{\mathbf{a}}]_{m+\frac{M}{2}} $ can be caused by either $ b_{hm}=-1 $ or $ b_{h(m+\frac{M}{2})}=-1 $, incurring ambiguity in RSR detection.
To remove the ambiguity, we need to enforce a protocol between radar and Bob that RSR only happens on the antenna associated with smaller (or larger) \HFs. This constraint can be expressed as
\begin{align}\label{eq: C2}
b_{hm^*}=-1~~\mathrm{s.t.}~m^*:~\min_{m,m+M/2}\{ k_{hm},k_{h(m+M/2)}  \}.
\end{align}

\begin{algorithm}[!t] %
	
	\caption{Removing RSR for Bob}
	\label{alg: detect RSRs}
	\begin{algorithmic}[1]	
		
		\State{\textbf{Input}: $ Q $, $ M $ and $ \breve{\mathbf{y}}_h $;} \label{stepA2: input}
		\Comment{{\footnotesize $ |\cdot|^2 $ takes element-wise absolute square}}
		\State{Calculate $ \mathbf{Y}_h = |\breve{\mathbf{y}}_h\mathbf{1}^{\mathrm{T}} - \mathbf{1}\breve{\mathbf{y}}_h^{\mathrm{T}}|^2 $;}\label{stepA2: Yh=yh-yh'}
		
		\State{Identify the minimum element (excluding diagonal element) and its index in each row of $ \mathbf{Y}_h $, stacking them in $ \mathbf{y}_{\mathrm{min}} $ and $ \mathbf{d} $, respectively;}\label{stepA2: identify min and index}
		
		\State{Sort $ \mathbf{y}_{\mathrm{min}} $ in ascending order and denote the index vector from $ \mathbf{y}_{\mathrm{min}} $ to the sorted version as $ {\mathbf{d}}_1 $;} \label{stepA2: sort min}
		
		\For{$ q=0:Q-1 $}	\label{stepAlg1: for q=}
		\Comment{{\footnotesize Index starts from $ 0 $.}}
			\State{Take $ i=[{\mathbf{d}}_1]_q $ and $ i'=[\mathbf{d}]_i $. Remove $ i' $ from $ {\mathbf{d}}_1 $;} \label{stepA2: get indexes for comparison}
			
			\State{$ [\mathbf{y}_{h}]_{i^*}=-[\mathbf{y}_{h}]_{i^*} $, s.t. $ i^*=\min_{i,i'}\Big\{\frac{L-l_{i}^*}{\mn},\frac{L-l_{i'}^*}{\mn}\Big\}$;} \label{stepA2: remove RSR}
		\EndFor  \label{stepA2: end q=}
		
		\State{\textbf{Return} $ \tilde{\mathbf{y}}_h=\breve{\mathbf{y}}_h $.}
		
	\end{algorithmic}
\end{algorithm}

Algorithm \ref{alg: detect RSRs} is designed to  remove RSR for Bob based on $ \breve{\mathbf{y}}_h $ given in (\ref{eq: bf y_h rewritten}). In Step \ref{stepA2: Yh=yh-yh'}, the power differences between each element in $ \mathbf{y}_h $ and all the other elements are calculated.
After Steps \ref{stepA2: identify min and index} and \ref{stepA2: sort min}, the first several elements in $ {\mathbf{d}}_1 $ are related to the indexes of RSR antennas. In Step \ref{stepA2: get indexes for comparison}, the indexes of the two antennas, whose received signals are most similar in power, are extracted. By comparing the associated \HFs, RSR is detected with the aid of constraint (\ref{eq: C2}) and removed by reversing back the sign; see Step \ref{stepA2: remove RSR}. 
{It is noteworthy that the removal of RSR at Bob also owes to EPC that only works for Bob (due to the AoD-specific design of EPC).
This indicates that Eve cannot remove RSR as Bob does in Algorithm \ref{alg: detect RSRs}.}  

\subsection{Impact of Proposed Design on Radar Performance}\label{subsec: impact on radar}
Subsequently, we illustrate the impact of each module in the proposed baseband waveform processing on radar performance using the range ambiguity function. Consider an FH-MIMO radar with $ M $ antennas and $ H $ hops per pulse. Let $ \tau $ denote time delay.  
Based on \cite[Eq. (27)]{Amb_FH_MIMO2008TSP}, we can express the range ambiguity function of the radar as,
\begin{align}\label{eq: chi(tau)}
R(\tau) = \left| \sum_{m=0}^{M-1}\sum_{m'=0}^{M-1}\sum_{h,h'=0}^{H-1} \underbrace{\chi(\tilde{\tau},{\nu})e^{\mj2\pi {\nu}{hT}}}_{\mathcal{B}} \underbrace{e^{\mj2\pi f_{h'm'}\tau}}_{\mathcal{D}} \right|,
\end{align}
where $ \tilde{\tau} = \tau-{T(h'-h)} $, $ {\nu}=f_{hm} - f_{h'm'} $ and $ \chi(x,y) $ is the ambiguity function of a standard rectangular pulse with $ x $ and $ y $ spanning range and Doppler domains, respectively. 
According to \cite[Eq. (26)]{Amb_FH_MIMO2008TSP}, we have 
\begin{align}
\chi(x,y) = \Big(T-|x|\Big)\mathcal{S}\Big(y\big(T-|x|\big)\Big)e^{\mj\pi y(x+T)},~\mathrm{if}~|x|<T;\nonumber
\end{align}
and otherwise $ \chi(x,y)=0 $, where $ \mathcal{S}(\alpha)=\frac{\sin(\pi\alpha)}{\pi\alpha} $. The impact of proposed processing on $ R(\tau) $ is analyzed below.

\subsubsection{Impact of HFCS on $ R(\tau) $}
HFCS selects $ M $ out of $ K $ \textit{different} \HF~per hop based on varying information bits to be transmitted. The waveform orthogonality condition given in (\ref{eq: wavefor orthogonality}) is hence always satisfied under HFCS processing. 
As conventional FH-MIMO radars randomly selects \HFs~\cite{Amb_FH_MIMO2008TSP}, HFCS, resembling the random selection, incurs negligible changes to the key features of $ R(\tau) $, e.g., mainlobe width and mainlobe-to-sidelobe ratio etc.  

\subsubsection{Impact of HFPS on $ R(\tau) $}
We see from (\ref{eq: chi(tau)}) that $ R(\tau) $ is determined by the combinations of $ (\mathcal{B},\mathcal{D}) $ which is in essence relied on the combinations of $ (\nu,f_{h'm'}) $. By fixing $ f_{h'm'} $, the combinations of $ (\nu,f_{h'm'}) $ remain the same despite the ordering of the \HFs~at hop $ h $. The same conclusion holds by fixing $ f_{hm} $ and randomly changing the ordering of the \HFs~at hop $ h' $. 
This is validated by the example given in Tables \ref{tab: FH sequences} and \ref{tab: combinations of (nu,fph m')}, 
where, C1 and C2 in Table \ref{tab: FH sequences} give two orderings of the same \HFs, and, clearly, the overall combination set of $ \big({\nu},f_{h'm'}\big) $ obtained under C1 is identical to that of $ \big(\tilde{\nu},\tilde{f}_{h'm'}\big) $ under C2. Therefore, we can claim that HFPS does not incur any change to $ R(\tau) $ after \HFs~are selected by HFCS.

\begin{table}[!t]
	\caption{Different FH Sequences}
	\label{tab: FH sequences}
	\vspace{-10pt}
	\begin{center}
		\begin{tabular}{c|c|c|c|c|c}
			\hline
			& $ (h,m) $ & $ (0,0) $ & $ (0,1) $ & $ (1,0) $ & (1,1) \\\hline
			C1& $ f_{hm} $ (MHz)	& $ 20 $ & $ 10 $ & $ 45 $ & $ 30 $\\\hline
			C2& $ \tilde{f}_{hm} $ (MHz)	& $ 10 $ & $ 20 $ & $ 30 $ & $ 45 $\\\hline
		\end{tabular}
	\end{center}
\end{table}
\begin{table}[!t]
	\caption{Combinations of $ \big({\nu},f_{h'm'}\big) $, where the frequency is in MHz}
	\label{tab: combinations of (nu,fph m')}
	\vspace{-10pt}
	\begin{center}
		\begin{tabular}{c|c|c|c|c}
			\hline
			$ {(m,m')} $& $ (0,0) $ & $ (0,1) $ & $ (1,0) $ & $ (1,1) $\\\hline
			$ {(h,h')} $ & \multicolumn{4}{c}{$ (0,1) $} \\\hline
			$ \big({\nu},f_{h'm'}\big) $, C1	& $ (-20,30) $ & $ (-35,45) $ & $ (-10,30) $ & $ (-25,45) $\\\hline
			$ \big(\tilde{\nu},\tilde{f}_{h'm'}\big) $, C2	& $ (-25,45) $ & $ (-10,30) $ & $ (-35,45) $ & $ (-20,30) $\\\hline
		\end{tabular}
	\end{center}
\end{table}

\subsubsection{Impact of EPC and RSR on $ R(\tau) $}
According to (\ref{eq: EPC waveform}) and (\ref{eq: bf y_h rewritten}), the joint impact of EPC and RSR is that the phases of radar-transmitted signals are randomly modulated across antennas and hops. 
As analyzed in \cite{DFRC_AmbiguityFunc2018Amin}, PSK modulations can prevent periodic coherent accumulation (which occurs whenever $ \tau $ is integer times of a hop duration), hence suppressing periodic sidelobe spikes of $ R_\tau $. Given the equivalence between the impact of EPC and RSR on radar signals and that of PSK \cite{DFRC_AmbiguityFunc2018Amin}, we conclude that EPC and RSR can suppress periodic sidelobe spikes of $ R(\tau) $. 
A benefit of the suppression is the reduced mutual interference among radar targets.
This will be validated in Section \ref{sec: simulation}.

\section{Numerical Illustration of the Proposed Design}\label{sec: numerical running proposed algorithms}

Having introduced each module in the proposed waveform processing (refer to Fig. \ref{fig: system model}), we provide a numerical example to demonstrate the overall workflow of incorporating the proposed design in an FH-MIMO DFRC system. 
For illustration convenience but without loss of generality, small values of parameters are taken: $ M=4 $, $ K=5 $ and $ Q=1 $. The task here is: \textit{transmit symbol $ \mathbf{e}=[01~0011]^{\mathrm{T}} $ to Bob at hop $ h $}.

\textit{I) Initialization:} According to Section \ref{subsec: overall scheme of the proposed waveform processing}, the number of bits able to be conveyed by HFCS is $ E_1=\lfloor \log_2C_M^K \rfloor=2 $ and that by HFPS is $ E_2=\lfloor \log_2(M!) \rfloor=4 $. Thus, each radar hop can transmit $ E=E_1+E_2=6 $ bits.
Out of $ C_M^K=5 $ different combinations, $ 2^{E_1}=4 $ combinations are selected as HFCS constellation points and collected by $ \mathcal{C}_1 $. We take $ \mathcal{C}_1 $ as 
\begin{align}
\left\{\begin{array}{c}
[\mathcal{C}_1]_0=\{0,1,2,3\},~[\mathcal{C}_1]_1=\{0,1,2,4\},\\
{}[\mathcal{C}_1]_2=\{0,1,3,4\},~[\mathcal{C}_1]_3=\{0,2,3,4\}
\end{array}\right\},\nonumber
\end{align}
where $ [\mathcal{C}_1]_i $ denotes the $ i $-th constellation point in $ \mathcal{C}_1 $. 
Out of $ M!=16 $ permutations of \HFs, $ 2^{E_2}=16 $ permutations are selected as HFPS constellation points and collected by $ \mathcal{C}_2 $, as given by
\begin{align}
	\left\{\begin{array}{c}
		[\mathcal{C}_2]_0=\{3,2,1,0\},~[\mathcal{C}_2]_1=\{3,2,0,1\}, \\
		{}[\mathcal{C}_2]_2=\{3,1,2,0\},~[\mathcal{C}_2]_3=\{3,1,0,2\} ,\\
		\cdots\cdots
	\end{array}\right\}.\nonumber
\end{align}

\textit{II) HFCS and HFPS:} The first $ E_1(=2) $ bits of $ \mathbf{e} $ are used to perform HFCS. 
Since $ (01)_{\mathrm{D}}=1 $, $ [\mathcal{C}_1]_1=\{0,1,2,4\} $ is selected as the set of \HFs, i.e., the zero-th, first, second and fourth sub-bands are used for radar transmission at hop $ h $.
Afterwards, $ (\cdot)_{\mathrm{D}} $ gives the decimal value of the enclosed bit sequence.
Then, the last $ E_2(=4) $ bits of $ \mathbf{e} $ are used for HFPS. 
Since $ (0011)_{\mathrm{D}}=3 $, $ [\mathcal{C}_2]_3 =\{3,1,0,2\}$ is selected to pair \HFs~with antennas.
The first element in $ [\mathcal{C}_2]_3 $ is $ 3 $, which indicates that the $ 3 $rd element in $ [\mathcal{C}_1]_1 $ is used for antenna $ m=0 $, i.e., $ k_{h0}=4 $. (Here, index starts from $ 0 $.) 
Accordingly, we obtain $ \mathbf{k}_h=[4,1,0,2]^{\mathrm{T}} $. 

\textit{III) EPC:} Substituting $ {k}_{hm}~\forall m $ into (\ref{eq: transmitted signal p h m}) gives $ s_{hm}(t) $. Then, further substituting $ s_{hm}(t) $ and $ \phi $ into (\ref{eq: EPC waveform}), the EPC-processed waveform is obtain, i.e., $ \tilde{s}_{hm}(t) $.

\textit{IV) RSR:} Initialize the coefficients caused by RSR as $ b_{hm}=1~\forall m $.
Given $ Q=1 $, a random integer is generated to be the index of RSR antenna. Take $ I_0=0 $ for the index. Enforcing constraint (\ref{eq: C2}), we set $ b_{h(I_0+M/2)}=-1 $, since $ k_{hI_0} > k_{h(I_0+M/2)} $. By multiplying $ b_{hm}  $ to $ \tilde{s}_{hm}(t)   $, RSR-processed signal $ \breve{s}_{hm}(t)=b_{hm}\tilde{s}_{hm}(t) $ is obtained. Then, radar radiates $ \breve{s}_{hm}(t) $ from antenna $ m $ in RF band.

\begin{figure}[!t]
	\centerline{\includegraphics[width=80mm]{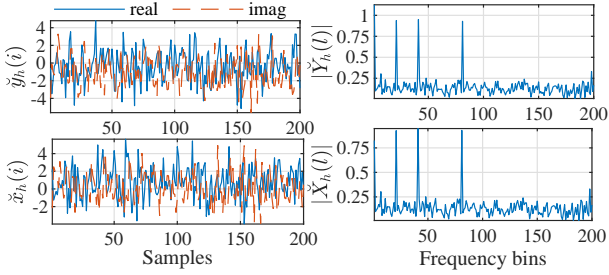}}
	\caption{Signals received by Bob and Eve in time and frequency domains.}
	\label{fig: Bob and Eve received signals}
	\vspace{-10pt}
\end{figure}

\textit{V) HFCS Decoding at Bob:} The baseband signal received at Bob, i.e., $ \breve{y}_h(i) $, is shown in Fig. \ref{fig: Bob and Eve received signals}. We see that the useful signal is corrupted in noises in the time domain. 
Taking a $ 200 $-point DFT leads to the frequency-domain signal as done in (\ref{eq: DFT of communciation signal}), we obtain $ \breve{Y}_h(l) $, whose ammplitude is shown in Fig. \ref{fig: Bob and Eve received signals}\footnote{Note that the time- and frequency-domain signals received by Eve are also provided in Fig. \ref{fig: Bob and Eve received signals} for comparison. We clearly see four peaks in the frequency spectrum of Eve and the peaks are located at the same discrete frequencies as those of Bob. Thus, Eve detects the same \HFs~as Bob.}. We see four peaks of $ |\breve{Y}_h(l)| $ and their indexes are $ l_0^*=0, ~l_1^*=20,~l_2^*=40 $ and $ l_3^*=80 $. 
Substituting $ l_m^* $ into (\ref{eq: Kh}), we obtain $ \widehat{\mathcal{K}}_h=\{0,1,2,4\} $, where $ \hat{x} $ denotes an estimate of $ x $.
Looking up $ \widehat{\mathcal{K}}_h $ in $ \mathcal{C}_1 $ gives its index in the set which is one in this example. 
Converting one to $ E_1 $ bits, the HFCS sub-symbol is decoded as ``$ 01 $''.

\textit{VI) RSR Removal at Bob:} Next, we perform Algorithm \ref{alg: detect RSRs} to remove RSR for Bob. Collecting $ \breve{Y}_h(l_m^*) $ for $ m=0,1,\cdots,4 $, we obtain $ \breve{\mathbf{y}}_h=[-0.3416 - 1.0724\mj,
-0.9379 + 0.0325\mj, 0.9413 - 0.1311\mj,
0.0198 - 0.9291\mj]^{\mathrm{T}} $ with AWGNs added.
Running Step \ref{stepA2: Yh=yh-yh'} of Algorithm \ref{alg: detect RSRs} leads to 
\begin{align}
	\breve{\mathbf{Y}}_h = \left[ \begin{array}{cccc}
	0 & 1.5764 & 2.5319 & 0.1511 \\
	1.5764 & 0 & 3.5584 & 1.8420 \\
	2.5319 & 2.5584 & 0 & 1.4860 \\
	0.1511 & 1.8420 & 1.4860 & 0
	\end{array} \right]	\nonumber
\end{align} 
Then Step \ref{stepA2: identify min and index} gives $ \mathbf{y}_{\mathrm{min}}=[0.1511,    1.5764,    1.4860 ,   0.1511]^{\mathrm{T}} $ and $ \mathbf{d}=[3,0,3,0]^{\mathrm{T}} $. Sorting $ \mathbf{y}_{\mathrm{min}} $ in Step \ref{stepA2: sort min} leads to $ \mathbf{d}_1=[0,3,2,1]^{\mathrm{T}} $. 
This further results in $ i=0 $ and $ i'=3 $ in Step \ref{stepA2: get indexes for comparison} of Algorithm \ref{alg: detect RSRs}. By comparing $ l_0/\mn $ and $ l_3/\mn $ in Step \ref{stepA2: remove RSR}, RSR is detected on $ [\breve{\mathbf{y}}_h]_0 $ and removed by reversing the sign of $ [\breve{\mathbf{y}}_h]_0 $.
The output of Algorithm \ref{alg: detect RSRs} is the RSR-removed signal, i.e., $ \tilde{\mathbf{y}}_h=[0.3416 + 1.0724\mj,
-0.9379 + 0.0325\mj, 0.9413 - 0.1311\mj,
0.0198 - 0.9291\mj]^{\mathrm{T}}. $

\textit{VII) HFPS Decoding at Bob:} 
With $ \tilde{\mathbf{y}}_h $ obtained, Algorithm \ref{alg: detect HFPS} is performed to decode HFPS sub-symbol. Substituting $ \tilde{\mathbf{y}}_h $ in Step \ref{stepA1: angle taking} gives $ \vec{\omega} = [3.0695,
-1.4878,
1.5972,
0.1131]^{\mathrm{T}}. $
In Step \ref{stepA1: angle revision}, the angles in $ \vec{\omega} $ are revised, leading to 
\begin{align}
\vec{\omega} = [-3.2137,
-1.4878,
-4.6859,
-6.1701]^{\mathrm{T}}.\nonumber
\end{align}
Sorting the revised $ \vec{\omega} $ gives the index vector $ \mathbf{i}=[1,0,2,3]^{\mathrm{T}}. $
Comparing $ \|e^{-\mj 6.1701}-1\|_2 =0.1130$ and $ \|e^{-\mj 1.4878}-1\|_2 =1.3543$ in Step \ref{stepA1: circshift}, we know that zigzag has affected $ \vec{\omega} $. Thus, $ \mathbf{i} $ is circularly shifted once, which gives $ \mathbf{i}=[3,1,0,2]^{\mathrm{T}}. $
Step \ref{stepA1: Ph construction} constructs $ \mathbf{P}_h $ based on $ \mathbf{i} $, as given by
\begin{align}
{\mathbf{P}}_h = \left[ \begin{array}{cccc}
0 & 0 & 1 & 0 \\
0 & 1 & 0 & 0 \\
0 & 0 & 0 & 1 \\
1 & 0 & 0 & 0
\end{array} \right].	\nonumber
\end{align} 
Substituting $ {\mathbf{P}}_h $ in (\ref{eq: hat k_h}), we obtain $ \hat{\mathbf{k}}_h=[4,1,0,2]^{\mathrm{T}} $.
Looking up $ \hat{\mathbf{k}}_h $ in $ \mathcal{C}_2 $ gives its index in the set which is three in this example. 
Converting decimal $ 3 $ to $ E_2(=4) $ bits, the HFPS sub-symbol is decoded as ``$ 0011 $''.
Both sub-symbols are correctly decoded at Bob applying the proposed methods.

\section{Simulation Results}\label{sec: simulation}
In this section, simulation results are presented to validate the proposed design. Unless otherwise specified, the FH-MIMO radar is configured as: $ M=4 $, $ Q=\frac{M}{2} $, $ K=20 $, $ H=15 $, $ B=100 $ MHz, $ T=1 ~\mu$s and $ L=200 $ (based on the sampling frequency of $ 2B $); and the communication parameters are: $ \phi \sim\mathcal{U}_{[-90^{\circ},90^{\circ}]} $, $ \theta\sim\mathcal{U}_{[-90^{\circ},90^{\circ}]} $, $ \alpha=e^{\mj x}~(x\sim\mathcal{U}_{[0,2\pi]})$ and $\beta=e^{\mj y} ~ (y\sim\mathcal{U}_{[0,2\pi]}) $.   
Here, $ \mathcal{U}_{[\cdot,\cdot]} $ stands for the uniform distribution in the subscript region.
Throughout simulation, Eve knows the AoD of Bob, if not otherwise specified.  
The time-domain SNR at Bob is defined based on (\ref{eq: comm-received signal}), as given by $ \gamma_{\mathrm{B}} =  \frac{M|\beta|^2}{\sigma_{\xi}^2} $, where $ \sigma_{\xi}^2 $ is the noise power of $ \xi(i) $. Based on (\ref{eq: bf y_h}), the decoding SNR at Bob is $ L\gamma_{\mathrm{B}} $, where the $ L $ times improvement is brought by DFT; see (\ref{eq: DFT of communciation signal}). Likewise, the time-domain and decoding SNRs at Eve are given by $ \gamma_{\mathrm{E}}=\frac{M|\alpha|^2}{\sigma_{z}^2} $ and $ L\gamma_{\mathrm{E}} $, respectively, where $ \sigma_{z}^2/L $ is the noise power of $ Z(l) $ given in (\ref{eq: DFT eave}). When presenting decoding performance, we use $E_b/N_0$, defined as energy per bit to noise power density ratio, i.e.,
\begin{align}\label{eq: Eb/n0}
	E_b/N_0 = L\gamma_{\mathrm{B}}BT/E,
\end{align}
where $ E $ is the number of bits conveyed per radar hop and $ \gamma_{\mathrm{B}} $ can be replaced with $ \gamma_{\mathrm{E}} $ to obtain $E_b/N_0$ for Eve.

 The labels used in the figures are interpreted as follows,
\begin{itemize}
	\item ``Bob-proposed'': indicates that EPC and RSR are performed at radar, and Algorithms \ref{alg: detect HFPS} and \ref{alg: detect RSRs} at Bob;
	
	\item ``Bob-without RSR'': indicates that EPC and Algorithm \ref{alg: detect HFPS} are performed at radar and Bob, respectively. This is the lower bound of ``Bob-proposed'' given the absence of RSR removal error caused by running Algorithm \ref{alg: detect RSRs};
	
	\item ``Eve'': indicates that EPC and RSR are performed at radar and (\ref{eq: ML eave}) is solved for HFPS decoding at Eve;
	
	\item ``Eve-without Bob's AoD'': is the same as above except the AoD of Bob is unavailable at Eve;
	
	\item ``Eve-without RSR'': is the same as ``Eve'' except RSR is not performed at radar. This also acts as a performance indicator of a general HFPS decoding without conducting the proposed EPC and RSR at radar;
	
	\item ``Bob/Eve-HFCS'': indicates that only HFCS is used for FH-MIMO DFRC, as done in the state of the art \cite{DFRC_FHcodeSel2018}.
\end{itemize}

\begin{figure}[!t]
	\centerline{\includegraphics[width=80mm]{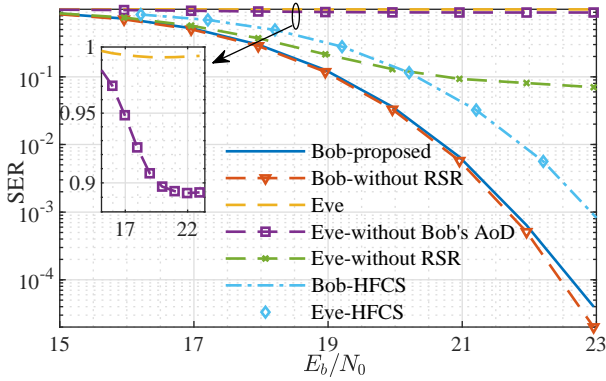}}
	\caption{SER against $E_b/N_0$, where $ \gamma_{\mathrm{B}} $ and $ \gamma_{\mathrm{E}} $ are both in $ [-10,-2] $ dB. The radar configuration leads to $ \lfloor \log_2C_K^M \rfloor= 12 $ bits conveyed by HFCS sub-symbol and $ \lfloor \log_2M! \rfloor=4 $ bits by HFPS. Substituting $ E=12+4 $ into (\ref{eq: Eb/n0}) gives the $E_b/N_0$ region in the figure. The zoomed-in sub-figure helps see the SER of Eve more clearly. 
	}
	\label{fig: SER vs snr}
\end{figure}

Fig. \ref{fig: SER vs snr} compares the SERs achieved by Bob and Eve as $E_b/N_0$ increases. 
From the curves labeled ``Bob-proposed'' and ``Eve'' in Fig. \ref{fig: SER vs snr}, we see that Bob has a decreasing SER against $E_b/N_0$ and Eve has a close-to-one SER over the same region of $E_b/N_0$. This demonstrates the substantially high communication secrecy achieved by the proposed design. 
Comparing the curves labeled ``Bob-proposed'' and ``HFCS'' in Fig. \ref{fig: SER vs snr}, we see the improvement of the proposed scheme over HFCS \cite{DFRC_FHcodeSel2018}. In particular, the proposed design reduces SER by more than one order of magnitude at $E_b/N_0$$ =23 $ dB.
This improvement owes to: (i) the use of HFPS which increases the number of bits conveyed per radar hop; (ii) the proposed EPC which solves the AoD-dependent issue of HFPS decoding; see the curves ``Bob-proposed'' (with EPC) versus ``Eve-without RSR'' (suffering from the issue); and (iii) the newly designed Algorithm \ref{alg: detect RSRs} which accurately removes RSR for Bob, c.f., the almost overlapping curves labeled as ``Bob-proposed'' and ``Bob-without RSR''. 

We see from Fig. \ref{fig: SER vs snr} that solely using HFCS leads to identical SER performance of Bob and Eve, which highlights the necessity and significance of our design of enhancing physical layer security.
From the three curves related to Eve in Fig. \ref{fig: SER vs snr}, we see three levels of performance degradation at Eve incurred by the proposed design. \textit{First}, by introducing HFPS, the decoding of Eve suffers from an AoD-dependent issue, leading to the convergence of an SER close to $ 0.1 $. \textit{Second}, by introducing EPC, the decoding at Eve substantially degrades without the AoD of Eve, increasing the converging SER to about $ 0.9 $. \textit{Third}, with RSR performed, the HFPS decoding at Eve is completely incapacitated, incurring SER larger than $ 0.99 $ across the whole region of $E_b/N_0$.

\begin{figure}[!t]
	\centerline{\includegraphics[width=80mm]{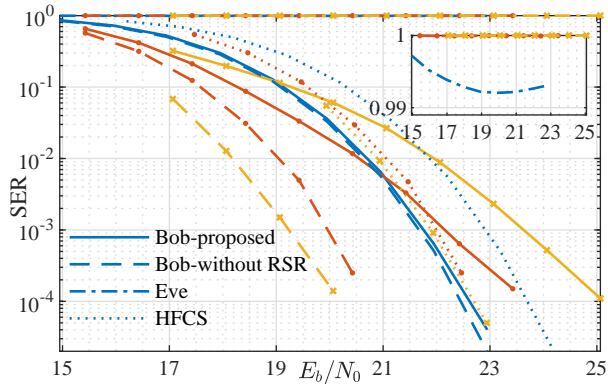}}
	\caption{SER against $E_b/N_0$, where $ M=4 $, $ 6 $ and $ 8 $ for curves without markers, with dots and with crosses, respectively, and $ Q=\frac{M}{2}~\forall M $. $ \gamma_{\mathrm{B}}\in[-10,-2] $ dB is set for $ M=4 $ and increased by $ 4 $ dB and $ 8 $ dB for $ M=6 $ and $ 8 $, respectively; the same for $ \gamma_{\mathrm{E}} $. The zoomed-in sub-figure helps see the SER of Eve more clearly.}
	\label{fig: SER vs snr diff M}
\end{figure}

Fig. \ref{fig: SER vs snr diff M} compares the SER performance achieved under different values of $ M $. 
Corresponding to $ M=4,6 $ and $ 8 $, the numbers of bits conveyed by HFPS are $ 4, 9 $ and $ 15 $, and those by HFCS are $ 12 $, $ 15 $ and $ 16 $, respectively. 
Comparing the curves labeled as ``Bob-without RSR'' and ``HFCS'', we see that the increased bits per hop earned by the proposed scheme has the potential of improving SER performance progressively as $ M $ increases. 
We also see from that the gap between the actual SER achieved by our design (``Bob-proposed'') and the bound (``Bob-without RSR'') increases with $ M $, rendering the proposed HFPS plus HFCS slightly worse than the sole HFCS in high $E_b/N_0$ regions. As expected, this is the price of the substantially high communication secrecy; see the zoomed-in sub-figures. It is noteworthy that, albeit the slight performance loss caused by Algorithm \ref{alg: detect RSRs}, the achievable SER sill improves with $E_b/N_0$, implying that the loss is compensable. In contrast, the converging SER at Eve is irrecoverable.

\begin{figure}[!t]
	\centerline{\includegraphics[width=80mm]{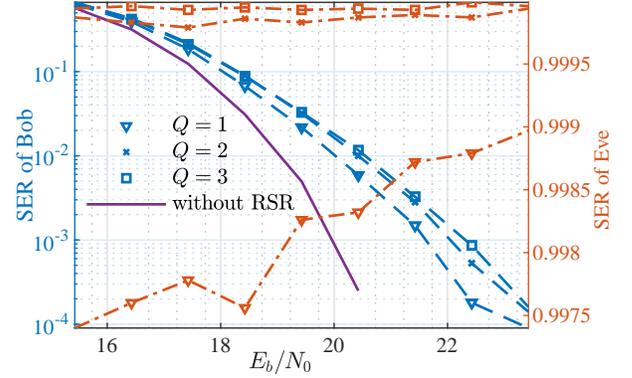}}
	\caption{SER achieved by ``Bob-proposed'' as $ {E}_{{b}}/N_0 $ increases, where $ M=6 $, $ Q $ takes $ 1 $ to $ 3 $, the dash-dotted curves use the $ y $-axis on the right, and both $ \gamma_{\mathrm{B}} $ and $ \gamma_{\mathrm{E}} $ are set in $ [-6,2] $ dB.}
	\label{fig: SER vs snr diff Q}
\end{figure}

Fig. \ref{fig: SER vs snr diff Q} observes the impact of $ Q $ on the proposed design. We see that the achievable SER increases negligibly with $ Q $. This validates the robustness of the newly designed Algorithm \ref{alg: detect RSRs} against $ Q $. We also see that the SER of Eve approaches one tightly even at $ Q=1 $ and is closer to one as $ Q $ increases. This validates our analysis in Appendix  \ref{app: fb Gaussian noise}; specifically, the SER convergence at Eve happens for sure, as the number of permutation matrices leading to error HFPS decoding
is larger than one at $ Q=1 $ and increases with $ Q $.

\begin{figure}[!t]
	\centerline{\includegraphics[width=80mm]{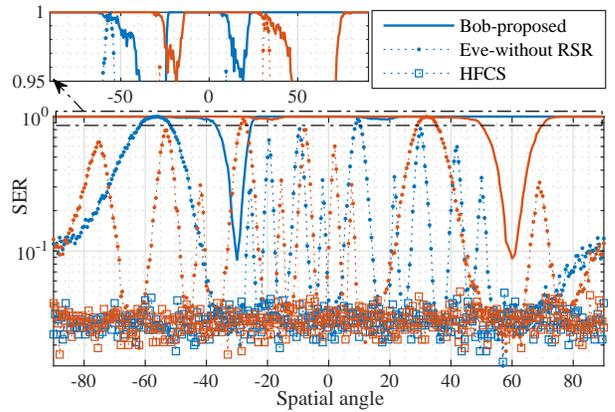}}
	\caption{SER against spatial angle from $ -90^{\circ} $ to $ 90^{\circ} $ with a grid of $ 0.5^{\circ} $, where $ \phi=-30^{\circ} $ and $ 60^{\circ} $ are observed, $ M=6 $, $ Q=3 $, and $ \gamma_{\mathrm{B}}=-3 $ dB. }
	\label{fig: SER vs angle}
		\vspace{-10pt}
\end{figure}

Fig. \ref{fig: SER vs angle} observes the SER performance against the spatial angle. From ``Bob-proposed'', we see that the proposed scheme can achieve high secrecy almost omnidirectionally. Except at the AoD of Bob, the SERs at all the other spatial angles approach one tightly. We also see that the SERs of Bob achieved at different AoDs are similar. This owes to the proposed EPC which removes the AoD-dependence for Bob. 
From the curve ``Eve-without RSR'', we see the reduced physical layer enhancement brought by EPC if the AoD of Bob is known to Eve. Moreover, from the curve ``HFCS'', we see that a uniform SER performance is achieved over the whole angular region. This again demonstrates that solely using HFCS for FH-MIMO DFRC is highly prone to eavesdropping.

\begin{figure}[!t]
	\centerline{\includegraphics[width=80mm]{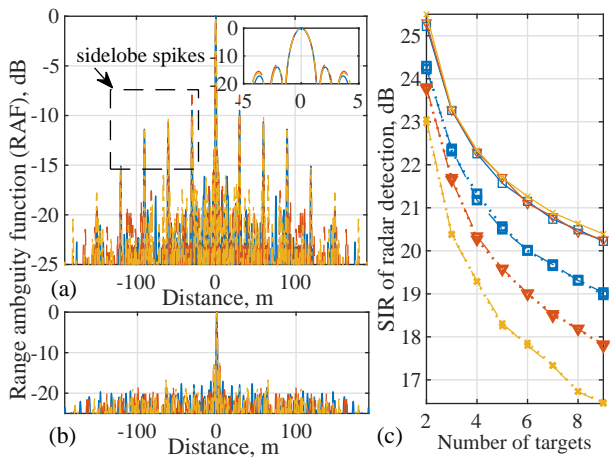}}
	\caption{Impact of the proposed baseband waveform processing on radar detection, where (a) RAFs under three groups of independently and randomly generated \HFs; (b) RAFs after performing HFPS, EPC and RSR on the waveform generated for Fig. \ref{fig: radar ambiguity}(a); and (c) SIR against number of targets. Figs. \ref{fig: radar ambiguity}(a) and \ref{fig: radar ambiguity}(b) take $ M=8 $. In \ref{fig: radar ambiguity}(c), dotted curves are based on conventional FH-MIMO radar waveform with randomly generated \HFs, dash curves are based on HFCS, and solid curves are based on all the proposed processing; and square, triangle and cross markers correspond to $ M=4,6 $ and $ 8 $, respectively.}
	\label{fig: radar ambiguity}
\end{figure}

Last but not least, we illustrate the impact of the proposed waveform processing on FH-MIMO radar performance. 
Fig. \ref{fig: radar ambiguity}(a) is provided to illustrate the impact of HFCS on radar ranging performance. We see that the mainlobes and mainlobe-to-sidelobe ratios (MSRs) under three realizations are almost identical. We also see that the periodicity of sidelobe spikes is the same for different random sets of \HFs. From Fig. \ref{fig: radar ambiguity}(b), we see that the proposed waveform processing suppresses the sidelobe spikes, improving the minimum MSR by more than $ 10 $ dB. 
From Fig. \ref{fig: radar ambiguity}(c), we see that the improvement on MSR increases SIR of radar detection by up to $ 4 $ dB at $ M=8 $. We also see that, despite the values of $ M $, the proposed processing leads to similar SIRs. We further see that HFCS waveform achieves the same SIR as the conventional FH-MIMO radar waveform. 
These observations of Fig. \ref{fig: radar ambiguity} validate our analysis in Section \ref{subsec: impact on radar}.

\section{Conclusion}
In this paper, a secure and high-speed FH-MIMO DFRC system is developed. This is achieved by introducing HFPS constellations to fully exploit information embedding capability embodied in \HFs. This is also accomplished by a new EPC processing which addresses the AoD-dependent issue for Bob and substantially enhances the physical layer security if the AoD of Bob is unknown to Eve. This is further fulfilled by the proposed RSR which scrambles constellations almost omnidirectionally and forces the SER of Eve, even knowing the AoD of Bob, into converging to one. 
Validated by simulations, our new design achieves substantially high secrecy, increases data rate and improves SIR of radar detection. As a future work, we will introduce multi-antenna receiver for Bob and develop new methods to further improve decoding performance.

\appendix

\subsection{Calculation of (\ref{eq: DFT of communciation signal})}\label{app: derivations of Bob's DFT}

Based on (\ref{eq: comm-received signal}), the $ L $-point DFT of the signal component in $ y_h(i) $, denoted by $ Y_h^{\mathrm{s}}(l) $, can be calculated as
\begin{align}
& Y_h^{\mathrm{s}}(l) = \beta \sum_{m=0}^{M-1}e^{-\mj mu_{\phi}}  \frac{1}{L}\sum_{i=0}^{L-1} e^{\mj2\pi \mn ik_{hm}/L} e^{-\mj \frac{2\pi l i}{L}}
\nonumber\\
& 
= \beta \sum_{m=0}^{M-1}e^{-\mj mu_{\phi}}  \frac{\sin\left( \pi L\left(\frac{l-\mn k_{hm}}{L}\right) \right)}{L\sin\left( \pi \left(\frac{l- \mn k_{hm}}{L}\right) \right)}e^{-\mj 2\pi\frac{L-1}{2}\left(\frac{l- \mn k_{hm}}{L}\right)}.\nonumber
\end{align}
From the above result, we see that $ \sin\left( \pi L\left(\frac{l-\mn k_{hm}}{L}\right) \right) $ is always zero due to the integer $ \mn $; see (\ref{eq: wavefor orthogonality}). This indicates that $ Y_h^{\mathrm{s}}(l) $ is only non-zero when the denominator equals to zero, i.e., $ \sin\left( \pi \left(\frac{l- \mn k_{hm}}{L}\right) \right)=0 $. The sine function only takes zero provided $ \left(\frac{\mn k_{hm}+l}{L}\right) $ is an integer which leads to $ l=\mn k_{hm} $. 
From the above analysis, $ Y_h^{\mathrm{s}}(l) $ only takes non-zero values at $ l=\mn k_{hm} $. Using the Dirac function, $ Y_h^{\mathrm{s}}(l) $ is written as in (\ref{eq: DFT of communciation signal}). 

\subsection{Proof of Proposition \ref{pp: fb Gaussian noise and SER convegernce}}\label{app: fb Gaussian noise}

We first illustrate that $ {f}_{b}^{\Re}({\mathbf{P}}) $ conforms to a normal distribution.
Denote $ {\mathbf{P}}^{\mathrm{T}}{\mathbf{P}}_h $ in (\ref{eq: objective function change of ML eave}) as $ \breve{{\mathbf{P}}} $ and rewrite $ {f}_{b}^{\Re}({\mathbf{P}}) $ as 
\begin{align}
{f}_{b}^{\Re}({\mathbf{P}}) = \Re\left\{ {\mathbf{a}}_{\theta\phi}^{\mathrm{H}} \big(\breve{\mathbf{P}}\mathbf{b}_h\odot \breve{\mathbf{P}}{\mathbf{a}}_{\theta\phi}\big)  \right\}=\Re\left\{ \sum_{m=0}^{M-1} \breve{b}_{hm} e^{\mj \Delta_{hm}}  \right\},\nonumber
\end{align}
where $ \breve{b}_{hm} = [\breve{\mathbf{P}}\mathbf{b}_h]_m $ and $ \Delta_{hm} $ is the difference between the phases of $ [\mathbf{a}_{\theta\phi}]_m $ and $ [\breve{\mathbf{P}}{\mathbf{a}}_{\theta\phi}]_m $, i.e.,
\begin{align}\label{eq: Delta_hm}
	 \Delta_{hm} = m{(2\pi/M+{u}_{\theta\phi})} - \arg\left\{[\breve{\mathbf{P}}\mathbf{{a}}_{\theta\phi}]_m\right\}~\forall m, 
\end{align}
where $ \mathbf{{a}}_{\theta\phi} $ is given in (\ref{eq: ML eave}) and $ u_{\theta\phi}=u_{\theta}-u_{\phi} $.
Based on (\ref{eq: P bhm}), we have 
\begin{align}\label{eq: P breve b_hm}
\mathbb{P}\{\breve{b}_{hm}=1\}={(M-Q)}/{M};~\mathbb{P}\{\breve{b}_{hm}=-1\}={Q}/{M}.
\end{align}
As $ \breve{b}_{hm}~ (m=0,1,\cdots,M-1) $ are independent Bernoulli-like variables, 
a weighted sum of them with constant-modulus weights, i.e., $ \sum_{m=0}^{M-1} \breve{b}_{hm} e^{\mj \Delta_{hm}} $, approaches to a normally distributed variable according to the central limit theorem \cite{PLS_hybridArray_TVT_Robert}. 
As the real part of a complex normal variable, $ {f}_{b}^{\Re}({\mathbf{P}}) $ also conforms to a normal distribution, i.e., $ {f}_{b}^{\Re}({\mathbf{P}})\sim \mathcal{N}(\mu_f,\sigma_f^2) $.

We proceed to calculate the parameters of $ {f}_{b}^{\Re}({\mathbf{P}}) $. 
Taking the expectation of $ {f}_{b}^{\Re}({\mathbf{P}}) $ leads to
\begin{align} \label{eq: mu_f}
\mu_f & = \mathbb{E}\left\{ {f}_{b}^{\Re}({\mathbf{P}}) \right\}  = \Re\Bigg\{\sum_{m=0}^{M-1}\underbrace{\mathbb{E}\left\{    \breve{b}_{hm}\right\}}_{\mu_{f1}} \underbrace{\mathbb{E}\left\{e^{\mj \Delta_{hm}}\right\}}_{\mu_{f2}}\Bigg\} ,        
\end{align}
where the randomness in $ e^{\mj \Delta_{hm}} $ is caused by $ \breve{\mathbf{P}} $; see (\ref{eq: Delta_hm}). 
Based on (\ref{eq: P breve b_hm}), we can calculate $ \mu_{f1} $ as
\begin{align}\label{eq: mu f1}
	\mu_{f1} = \left(1\times \frac{M-Q}{M}+(-1)\times \frac{Q}{M}\right)=\frac{M-2Q}{M}.
\end{align}
Given a large sample set of $ \breve{{\mathbf{P}}} $, the second term on the RHS of (\ref{eq: Delta_hm}) has a uniformly distributed angle, i.e.,
\begin{align}
\mathbb{P}\left\{\arg\left\{[\breve{\mathbf{P}}\mathbf{{a}}_{\theta\phi}]_m\right\}=m'\big(u_{\theta\phi}+\frac{2\pi}{M}\big)\right\}=\frac{1}{M}~\forall m'\le M-1.\nonumber
\end{align}
Based on the above PDF and (\ref{eq: Delta_hm}), $ \mu_{f2} $ can be calculated as
\begin{align} \label{eq: mu f2}
	&\mu_{f2} = e^{\mj m\big(u_{\theta\phi}+\frac{2\pi}{M}\big)}\left(\sum_{m'=0}^{M-1} \frac{e^{-\mj m'\big(u_{\theta\phi}+\frac{2\pi}{M}\big)}}{M}\right)\\
	&=e^{\mj m\big(u_{\theta\phi}+\frac{2\pi}{M}\big)}  \times \frac{e^{-\mj\frac{M-1}{2}\big(u_{\theta\phi}+\frac{2\pi}{M}\big)}}{M} \frac{\sin\left(\frac{M\big(u_{\theta\phi}+\frac{2\pi}{M}\big)}{2}\right)}{\sin\left(\frac{\big(u_{\theta\phi}+\frac{2\pi}{M}\big)}{2}\right)}.\nonumber
\end{align} 
Substituting (\ref{eq: mu f1}) and (\ref{eq: mu f2})
into (\ref{eq: mu_f}) and after some manipulations, we obtain
\begin{align}\label{eq: mu_f final}
	\mu_f 
	=  \Re\left\{  \frac{M-2Q}{M^2} \frac{\sin^2\left(\frac{M\big(u_{\theta\phi}+\frac{2\pi}{M}\big)}{2}\right)}{\sin^2\left(\frac{\big(u_{\theta\phi}+\frac{2\pi}{M}\big)}{2}\right)} \right\},
\end{align}
which leads to (\ref{eq: mean and variance of artificial noise}).

\begin{figure*}[!t]%
	\begin{align}\label{eq: sigma^2_f}
	 \sigma_f^2  &=
	\frac{1}{2} \mathrm{var}\left\{ \sum_{m=0}^{M-1} \breve{b}_{hm} e^{-\mj \Delta_{hm}}  \right\} =
	\frac{1}{2}   \left\{ \mathbb{E} \left\{\left(\sum_{m=0}^{M-1} \left\{\breve{b}_{hm} e^{-\mj \Delta_{hm}}  \right\}\right)^2\right\} - \left(\mathbb{E} \left\{  \sum_{m=0}^{M-1} \breve{b}_{hm} e^{-\mj \Delta_{hm}}  \right\}\right)^2 \right\}
	\nonumber\\
	&\overset{(a)}{=}\frac{1}{2} \times \left( \left(\sum_{m=0}^{M-1} 1^2\times \frac{M-Q}{M}+(-1)^2\times \frac{Q}{M}\right)-\mu_f^2  \right)=\frac{1}{2}\times (M-\mu_f^2)
	\end{align}
\end{figure*}%
Next, we calculate the variance of $ {f}_{b}^{\Re}({\mathbf{P}}) $.
Given that the variance of the real part of a complex Gaussian-distributed variable is half the full variance, we calculate $ \sigma_f^2 $ as in (\ref{eq: sigma^2_f}), where $ \overset{(a)}{=} $ is achieved based on two facts. First, the expectations of products between different $ \breve{b}_{hm} $ are zero due to their mutual independence.
Second, $ \mathbb{E} \left\{  \sum_{m=0}^{M-1} \breve{b}_{hm} e^{-\mj \Delta_{hm}}  \right\}=\mu_f $ is plugged in, since $ \mathbb{E} \left\{  \sum_{m=0}^{M-1} \breve{b}_{hm} e^{-\mj \Delta_{hm}}  \right\} $, calculated within the curly brackets in (\ref{eq: mu_f final}), is already a real value.

We proceed to investigate the impact of the normally distributed $ {f}_{b}^{\Re}({\mathbf{P}}) $ on HFPS decoding at Eve. Referring to (\ref{eq: ML eave}), an error HFPS decoding happens at Eve when solving (\ref{eq: ML eave}) returns a $ \mathbf{P} $ satisfying the following event
\begin{align}\label{eq: event fb(P)>(M-2Q)}
	\mathcal{E}:~{f}_{b}^{\Re}({\mathbf{P}}) > {f}_{b}^{\Re}({\mathbf{P}}_h)=(M-2Q)~~\forall \mathbf{P}\ne \mathbf{P}_h.
\end{align}
Therefore, to prove that RSR causes the convergence of SER of Eve to one, we turn to validating there is always $ \mathbf{P} $ making event (\ref{eq: event fb(P)>(M-2Q)}) happen.

\begin{figure}[!t]
	\centerline{\includegraphics[width=80mm]{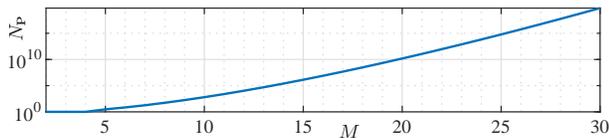}}
	\caption{Non-decreasing lower bound of $ N_{\mathbf{P}} $ derived in (\ref{eq: Np lower bound}) against $ M $.}
	\label{fig: Np against M}
\end{figure}

Given the normal distribution $ {f}_{b}^{\Re}({\mathbf{P}})\sim\mathcal{N}(\mu_f,\sigma_f^2) $, the probability of event $ \mathcal{E} $ can be expressed using the complementary error function, i.e.,
\begin{align}%
\mathbb{P}\{\mathcal{E}\} & = \frac{\mathrm{erfc}\left\{ h(\mu_f,Q) \right\}}{2}, \mathrm{s.t.}~h(\mu_f,Q)=\frac{(M-2Q)-\mu_f}{\sqrt{M-\mu_f^2}},\nonumber
\end{align}
where $ \sigma_f $ given in (\ref{eq: mean and variance of artificial noise}) has been plugged in. 
As illustrated in Appendix \ref{app: monotonicity of function against mu_f}, $ h(\mu_f,Q) $ is a non-increasing function of $ \mu_f\in[0,M-2Q] $. In addition, it can be readily validated that $ h(\mu_f,Q) $ is also a decreasing function of $ Q $. 
Therefore, given the decreasing monotonicity of $ \mathrm{erfc}\{\cdot\} $ against its argument, $ \mathbb{P}\{\mathcal{E}\} $ is maximized by taking $ \mu_f=0 $ and $ Q=1 $. This gives $ \mathbb{P}\{\mathcal{E}\} \ge  \frac{1}{2}\mathrm{erfc}\left\{ \frac{(M-2)}{\sqrt{M}} \right\}. $
Then the number of $ \mathbf{P} $ making $ \mathcal{E} $ happen, denoted by $ N_{\mathbf{P}} $, satisfies 
\begin{align}\label{eq: Np lower bound}
N_{\mathbf{P}} = \mathbb{P}\{\mathcal{E}\}\times M!\ge \frac{M!}{2}\mathrm{erfc}\left\{ \frac{(M-2)}{\sqrt{M}} \right\},
\end{align}
where $ M! $ is the number of all possible permutation matrices in $ \mathcal{P}_{\mathrm{Eve}} $; see (\ref{eq: ML eave}). As shown in Fig. \ref{fig: Np against M}, $ N_{\mathbf{P}} $ is a non-decreasing function of $ M $ and the lower bound of $ N_{\mathbf{P}} $ is one at the minimum $ M=2 $. This confirms the existence of $ \mathbf{P} $ causing (\ref{eq: event fb(P)>(M-2Q)}) for any values of $ M(\ge 2) $ and hence the convergence of the SER to one at Eve.

Before concluding the proof, we explain the angular region confinement in (\ref{eq: u_thetaPhi}). 
From (\ref{eq: sigma^2_f}), we see that to ensure a non-negative variance, $ M-\mu_f^2\ge 0 $ is required. By plugging (\ref{eq: mu_f final}) in the inequality, a region of $ u_{\theta\phi} $ can be obtained. Unfortunately, analytical expression for the region is unavailable, due to the discrete sinc function in (\ref{eq: mu_f final}). Nevertheless, it can be inferred that $ u_{\theta\phi} $ leading to $ M-\mu_f^2< 0 $ can only be located in the mainlobe of the sinc function in (\ref{eq: mu_f final}). This is because 
$ \mu_f^2 $ is upper bounded by $ \frac{16(M-2Q)^2}{81\pi^4} $ in the sidelobe regions\footnote{This is calculated by substituting $ u_{\theta\phi}=\frac{\pi}{M} $ into the sinc function in (\ref{eq: mu_f final}), since the peak of the first sidelobe is approximately achieved at the angle.}, i.e., $ 0\le  u_{\theta\phi}\le \frac{2\pi(M-2)}{M} $. 
For tractability, we use this region in Proposition \ref{pp: fb Gaussian noise and SER convegernce}, even though the actual region is slightly larger, as shown in Fig. \ref{fig: RSR mean and var}.

\subsection{Monotonicity of $ h(\mu_f,Q) $}\label{app: monotonicity of function against mu_f}
Taking the first partial derivative of $ h(\mu_f,Q) $ against $ \mu_f $, after lengthy yet straightforward manipulations, leads to 
\begin{align}
	\frac{\partial h(\mu_f,Q)}{\partial \mu_f} = \frac{\mu_f(M-2Q)-M }{(M-\mu_f^2)^{\frac{3}{2}}}.\nonumber
\end{align}
As $ \sigma_f^2 $ given in (\ref{eq: sigma^2_f}) is non-negative, we have $ M\ge \mu_f^2 $. Based on (\ref{eq: mu_f final}), we have $ \mu_f\le (M-2Q) $. Thus, the numerator of the above first partial derivative is non-positive. That is, $ h(\mu_f,Q) $ is a non-increasing function of $ \mu_f $.

\ifCLASSOPTIONcaptionsoff
  \newpage
\fi

\bibliographystyle{IEEEtran}
\bibliography{IEEEabrv,../ref/bib_JCAS.bib}

\end{document}